\documentclass[12pt]{article}

\usepackage{amssymb}
\newcommand{\C}{\mathbb{C}}
\newcommand{\R}{\mathbb{R}}
\newcommand{\Lieg}{\mathfrak{g}}

\title{\large A New Non-Perturbative Approach to Quantum Theory in Curved 
Spacetime Using the Wigner Function}
\author{Frank Antonsen\\ Niels Bohr Institute, Blegdamsvej 17\\ DK-2100
Copenhagen \O, Denmark}
\date{}
\begin{document}

\maketitle

\begin{abstract}
A new non-perturbative approach to quantum theory in curved
spacetime and to quantum gravity, based on a generalisation of the Wigner
equation, is proposed. 
Our definition for a Wigner equation differs from what have otherwise
been proposed, and does {\em not} imply any approximations. It is a completely
exact equation, fully equivalent to the Heisenberg equations of motion. The
approach makes different approximation schemes possible, e.g. it is possible to
perform a systematic calculation of the quantum effects order by order. An
iterative scheme for this is also proposed. 
The method is illustrated with some simple
examples and applications. A calculation of the trace of the renormalised
energy-momentum tensor is done, and the conformal anomaly is thereby related
to non-conservation of a current in $d=2$ dimensions and a relationship 
between a vector and an axial-vector current in $d=4$ dimensions.\\
The corresponding ``hydrodynamic equations'' governing the evolution of
macroscopic quantities are derived by taking appropriate moments.
The emphasis is put on the spin-$\frac{1}{2}$ case, but it is shown how to 
extend to arbitrary spins. Gravity is treated first in the Palatini formalism,
which is not very tractable, and then more successfully in the Ashtekar 
formalism, where the constraints lead to infinite order differential equations for
the Wigner functions.
\end{abstract}

\section*{Introduction}
One of the greatest problems we are facing in physics today is gravitation: as
is well-known we do not possess a well-behaved theory of quantum gravity. Even
semi-classical gravity, i.e. quantum field theory in curved spacetime, where
the gravitational field is kept as a classical background, is not that
well-behaved. It seems that perturbation theory is just not applicable to
problems involving gravitation, hence new, non-perturbative approaches have to
be found. In this paper, inspired by recent results in QCD (especially the
study of quark-gluon plasmas), we propose a non-perturbative approach based on
the Wigner operator. The equation governing the behaviour of this operator is
completely equivalent to the Heisenberg equations of motions for the fields.
This Wigner equation is the quantum analogue of the classical Boltzmann or
Vlasov equations, \cite{kinetic,van}, and thus allows for a better intuitive
understanding of the problems.\\
Start with a flat spacetime manifold and a Dirac spinor field, $\psi(x)$, 
on it.
Instead of just looking at $\psi(x)$ or its Fourier-transform $\tilde{\psi}(k)$
we consider a ``distribution'' $W(x,p)$ which is a functional of $\psi$ and
depends explicitly on both $x$ and $p$. This ``distribution'', the Wigner
operator, is, however, a non-positive operator, which is why we put the term
``distribution'' in quotation marks. The explicit form is given in the next
section; it is essentially a kind of Fourier transform of the density matrix. The
Dirac equation for $\psi$ gives rise to an integro-differential equation for
$W$, which is readily generalised to curved spacetimes. 
This allows us to attack the problem of fermions in curved space-time from
a new angle, and introduce new, non-perturbative, approximation schemes.\\
The next step is naturally to write down a similar equation governing the
gravitational fields. Since fermions causes torsion \cite{QFT,Ramond}, 
it would at first sight seem convenient to
introduce two Wigner operators $\Gamma_{\mu\nu}^{ab;cd}$ and $L_{\mu\nu}^{ab}$
for the spinor connection, $\omega_\mu^{ab}$, and the vier-bein, $e_\mu^a$,
respectively. These Wigner functions will not however be covariant, thus introducing
an unwanted gauge-dependency. In the Ashtekar formalism instead, we can introduce
two covariant Wigner functions, one for the $SU(2)$-connection, $A_i^a$, and one
for its conjugate $E^a_i$. In both cases, however, do the equations become highly
complicated, and we wont be able to say much about them.
We will basically, in this paper, consider the background
geometry as fixed (semi-classical quantum gravity). Let us note that such
an approach should find many applications: Hawking radiation, the early 
universe
etc. We will not, however, deal so much with these applications in this paper.\\
This is not the first time the powerful Wigner-equation methods have been
suggested as a framework for quantum fields in curved spacetime. Some earlier
work by Calzetta et al. and Kandrup, \cite{other}, uses either Riemann normal
coordinates
(and hence is only applicable in a sufficiently small region of spacetime)
or a mode-decomposition (which only works for static spacetimes), they do,
however, show the connection between this kinetic approach and the usual
path-integral or Green's function-based approach; the Wigner function can be 
expressed as a Fourier transform of a two point function with respect to the
middle point between the two points (the precise relationship is given in the next
section). In general, neither the Fourier transform nor the concept of a middle 
point makes sense in a general curved background, at least not in a coordinate
independent manner. Therefore one could attempt to use Riemann normal coordinates
in order to be able to define $x-x'$ for $x,x'$ points on the manifold, or one
could make use of a mode decomposition of the solutions to the equations of 
motion of a free field to generalize the Fourier transform. Neither of these two
approaches are completely general, although they do represent useful calculational
short-cuts. Winter, \cite{Winter}, has
proposed an approach similar to the one put forward here in which one integrates
along curves (geodesics, to be precise). But his approach needs the solution of
as well the geodesic as the geodesic deviation equations, a daunting task in
general. This again essentially restricts the usefulness of his method to
Riemann normal coordinates or to other approximation schemes (first quantum
corrections to the classical kinetic equation). The solution of the geodesic
equation is not needed for the approach presented here, even though we at an 
intermediate stage uses parallel transport along geodesics.\\
Halliwell, and
GellMann and Hartle has shown that the Wigner function appears naturally in
a quantum theory of histories, \cite{other}, as for instance quantum cosmology;
it is the only other place in which the Wigner function of the gravitational
degrees of freedom is studied.
It should finally be mentioned that Fonarev too, has extended the Wigner function
technique from flat space time to a curved background, \cite{Fonarev}. Like the
approach proposed here, he uses the structure of phase-space as a cotangent bundle,
our result, however, is much closer related to the analogy with Yang-Mills theory.
He furthermore emphasises the case of the scalar field, whereas this paper
mostly studies spinor fields, with some comments on the gravitational field. The
derivation of the conformal anomaly and of the ``hydrodynamic equations'' are also
new.\\
Most of the calculations are done in $d=4$ dimensions for concreteness but are valid
for all $d$. In some particular cases the case $d=2$ is also considered.

\section*{The Method}
We start out by considering the spacetime geometry as being fixed, i.e. we
begin by studying QFT in curved spacetime. Our main source of inspiration is
the papers by Vasak, Gyulassy and Elze \cite{EGV}, in which the Wigner operator
formalism is derived for Dirac fields interacting with Yang-Mills fields. We
will give a brief introduction to their results here, as this will indicate how
to generalise to curved spacetime.\\
Consider a second quantised Dirac field $\psi(x)$, the {\em Wigner operator} is
defined to be, \cite{van,EGV}
\begin{equation}
    \hat{W}(x,p) = \int e^{-iy\cdot p}\bar{\psi}(x+\frac{1}{2}y)\otimes\psi(x-
    \frac{1}{2}y)\frac{d^4y}{(2\pi)^4}
\end{equation}
note that $\hat{W}$ takes values in the Clifford algebra, a point we'll use
later. This definition can be made gauge invariant by replacing
\begin{equation}
    \psi(x-\frac{1}{2}y) = e^{-i\frac{1}{2}y\cdot\partial}\psi(x)
\end{equation}
by the gauge covariant generalisation
\begin{equation}
    \psi(x-\frac{1}{2}y)\equiv e^{-i\frac{1}{2}y\cdot\nabla}\psi
\end{equation}
where $\nabla$ is the (gauge) covariant derivative. Writing $x_\pm \equiv
x\pm\frac{1}{2}y$ we can write
\begin{equation}
    \hat{W}(x,p) =\int e^{-ip\cdot y}\bar{\psi}(x_+)U(x_+,x)\otimes U(x,x_-)
    \psi(x_-)\frac{d^4y}{(2\pi)^4}
\end{equation}
where
\begin{equation}
        U(b,a) \equiv P\exp\left(-ig\int_a^b A_\mu(z)dz^\mu\right)
\end{equation}
where $A_\mu$ denotes the Yang-Mills field, and the path of integration is a
straight line: $z(s) = a+(b-a)s$, which goes from $a$ to $b$ as $s$ goes from
$0$ to $1$. The Dirac equation then implies the following equation for
$\hat{W}$:
\begin{equation}
    \left[m-\gamma^\mu\left(p_\mu+\frac{1}{2}i\nabla_\mu\right)\right] \hat{W} =
    \frac{1}{2}ig\gamma^\mu\hat{X}_\mu\hat{W}
\end{equation}
where $\hat{X}_\mu$ is an integral operator involving the field strength tensor:
\begin{equation}
    \hat{X}_\mu\hat{W} \equiv -\frac{\partial}{\partial p_\nu}\left(\int_0^1
    (1-\frac{1}{2}s)e^{-\frac{i(1-s)}{2}\bigtriangleup}F_{\mu\nu}\hat{W}ds+
      \hat{W}
    \int_0^1\frac{1-s}{2}e^{\frac{1}{2}is\bigtriangleup}F_{\mu\nu}ds\right)
\end{equation}
where we have introduced the {\em triangle operator}
\begin{equation}
    \bigtriangleup \equiv \frac{\partial}{\partial p}\cdot\nabla
\end{equation}
It is understood that the derivative with respect to momentum always acts on the
Wigner distribution {\em alone} and {\em not} on any of the other terms.\\
One should also mention the relationship to Green's functions. If $G(x,x')$ denotes
the Green's function, $G(x,x') = \langle \bar{\psi}(x)\psi(x')\rangle$, then
the Wigner function can be rewritten as
\begin{displaymath}
	W(x,p) =\langle\hat{W}(x,p)\rangle 
	\propto \int e^{-iy\cdot p}G(x-\frac{1}{2}y,x+\frac{1}{2}y) d^dy
\end{displaymath}
i.e. as a Fourier transform of the two point function with respect to the distance
between the two points. This shows the relationship with the more familiar
approach to quantum theory using Green's functions.\\
The curvature 2-form $F_{\mu\nu}$ appears through the holonomy, i.e. through
re-expressing the derivative of the parallel transporter in the principal
bundle $U(b,a)$ in terms of a integration along a closed curve. 
The above equation actually holds 
for $A_\mu^k$ being an operator, but for a classical background one can pull
the field strength tensor out of the integral over $y$ to obtain
\begin{equation}
	\hat{X}_\mu = \pi^{-1/2}\frac{\partial}{\partial p_\nu} F_{\mu\nu}
	\left(
	j_0(\frac{1}{2}\bigtriangleup) -ij_1(\frac{1}{2}\bigtriangleup)
	\right)
\end{equation}
where $j_0,j_1$ are the usual spherical Bessel functions, $j_0\sim J_{1/2}, j_1\sim
J_{3/2}$. 
We will now analyse this equation and the steps leading to it.\\
First the obvious translations: the gauge field is replaced by the spinor
connection $\omega_\mu=\omega_\mu^{ab}\sigma_{ab}$, whereby the gauge covariant
derivative becomes the spinor covariant derivative, and the field strength
tensor becomes the Riemann-Christoffel curvature tensor $R_{\mu\nu}^{ab}$. 
Let us
next consider the path of integration. In flat space we integrate along the
straight line from $x_-=x-\frac{1}{2}y$ to $x_+=x+\frac{1}{2}y$, i.e.\ we go in
the direction $\pm y$ from $x$ for a ``period of time'' $\frac{1}{2}\|y\|$. The
curved space-time analogue is now obvious: $y$ is a tangent vector, and we
integrate along the (unique) geodesic with tangent $\pm y$ at $x$, the length of
the curve segment at each side of $x$ is $\frac{1}{2}\|y\|$. Note that this
simplifies things a lot: a priori we would have expected $y$ to lie in the
manifold, but now we see that it really lies in the tangent space. Hence for
each given point $x$ in the manifold we integrate over all possible tangents $y$
lying in the tangent space at $x$, $T_xM$, which is a flat space. This means
that we keep the measure $\frac{d^4y}{(2\pi)^4}$. To summarise: in flat 
space-time $x,y,p$ all belong to
the same space, whereas in curved space-time $x$ is in the manifold,  $y$ in
the tangent space and $p$ in the cotangent space. The dot-product $y\cdot p$ 
is then just the pairing between $T_xM$ and $T_x^*M$. What we have done is to make
extensive use of the intimate relationship between Yang-Mills theory as a theory of
connections on a principal bundle and general relativity as related to connections
on the frame bundle, \cite{GS,Ramond,Nash,Nakahara}.\\
We will follow customary notation and denote flat space indices by Roman letters
from the beginning of the alphabet, and curved spacetime ones by Greek letters. 
In curved spacetime, the Clifford
algebra relation $\{\gamma_\mu,\gamma_\nu\}=2g_{\mu\nu}$ implies that the Dirac
matrices are in general $x$-dependent, we introduce the vierbein as the
transformation connecting them to the usual flat (constant) Dirac matrices 
\cite{QFT,Ramond}:
\begin{equation}
    \gamma_\mu = e_\mu^a\gamma_a
\end{equation}
The connection between ``curved'' and ``flat'' indices is established by this
vierbein. The covariant derivative then reads, \cite{Ramond},
\begin{equation}
	\nabla_\mu = \partial_\mu +\omega_\mu^{ab}\sigma_{ab}
\end{equation}
with $\sigma_{ab}=\frac{1}{2}i[\gamma_a,\gamma_b]$ the generators of the
Lorentz algebra $so(3,1)$ in the spinor representation, and is the analogue of the
generator of the gauge algebra in the Yang-Mills case. The spin connection
$\omega_\mu^{ab}$ is related to the Christoffel symbol through
\begin{equation}
	\omega_\mu^{ab} = \eta^{ca}e^\nu_ce^b_\rho \Gamma^\rho_{\mu\nu}
	-\eta^{ca}e^b_\rho\partial_\mu e^\rho_c
\end{equation}
and it is this quantity which is the analogue of the Yang-Mills potential $A_\mu^a$.
Even though we deal with fermions, we will in general assume that no torsion
is present, i.e. $\Gamma^\rho_{\mu\nu}=\Gamma^\rho_{\nu\mu}$. This is a valid
assumption since torsion do not propagate and since we only want to deal with
fields in a given classical background.\\
With these comments the Wigner operator equation now reads
\begin{equation}
\left[m+\gamma^\mu\left(e_\mu^ap_a+\frac{1}{2}i\nabla_\mu\right)\right]\hat{W} =
-\frac{1}{2}\kappa\gamma^a\hat{X}_a\hat{W}
\end{equation}
where now
\begin{equation}
    \hat{X}_a\hat{W} \equiv \frac{1}{\sqrt{\pi}}
	e^\mu_a \frac{\partial}{\partial p_\nu}(j_0(\frac{1}{2}\bigtriangleup)
	-ij_1(\frac{1}{2}\bigtriangleup))R_{\mu\nu}^{bc}\{\sigma_{bc},\hat{W}\}
\end{equation}
with the triangle operator given by
\begin{equation}
    \bigtriangleup \equiv e^\mu_a\frac{\partial}{\partial p_a}\nabla_\mu
\end{equation}
The curly brackets denote an anticommutator. Both the curvature two-form
and the Wigner operator are Clifford algebra-valued, thus the Clifford algebra and not
only $so(3,1)$,
appears here as the analogue of the internal symmetry group of Yang-Mills theory.
One should remember that $so(3,1)\simeq spin(3,1)\subseteq C(3,1)$, so we have in a 
sense ``extended'' the gauge algebra from the Lorentz algebra to the full Clifford
algebra, with the price, of course, that we then no-longer have a Lie algebra but a
Clifford algebra instead. Also note that $so(3,1)\simeq spin(3,1)$ acts naturally on 
the Clifford algebra $C(3,1)$ through $x\mapsto [x,\sigma^{ab}]$, it is under this
Lie algebra that the Wigner function transforms. 
\\
Here we have just made the obvious translations from Yang-Mills theory to minimally
coupled Dirac fermions in a curved background. Non-minimally coupled fermions, i.e.
including a coupling to torsion, should be dealt with by making the substitution
$R_{\mu\nu}^{ab}\sigma_{ab} \rightarrow R_{\mu\nu}^{ab}\sigma_{ab} +
S^\rho_{\mu\nu}\nabla_\rho$, where $S^\rho_{\mu\nu}$ is the torsion (the
antisymmetric part of the Christoffel symbol). This is because the curvature only
enters through the commutator $[\nabla_\mu,\nabla_\nu] = 
R_{\mu\nu}^{ab}\sigma_{ab}$ 
(via the holonomy) and this gets modified to the above mentioned linear combination
of curvature and torsion when the latter is present, \cite{Ramond}. In a similar
fashion one could introduce gauge degrees of freedom, coupling to a Yang-Mills
field $A_\mu^k$ by adding $igA_\mu^kT_k$ to $\nabla_\mu$, where $T_k$ are the
generators of the gauged Lie algebra. We would then have to add $igF_{\mu\nu}^k
T_k$ to the right hand side of the commutator $[\nabla_\mu,\nabla_\nu]$. The
resulting Wigner function and its equation of motion would then also be gauge 
covariant.\\
Covariance is ensured by noting that $\psi$ transforms as $\psi\rightarrow {\cal U}
\psi$ where $\cal U$ is the transformation matrix (${\cal U} = \exp(i\alpha_{ab}(x)
\sigma^{ab})$ for the purely gravitational case and ${\cal U} = \exp(i\alpha_{ab}(x)
\sigma^{ab}+ig\alpha^k(x)T_k)$ with a coupling to a Yang-Mills field), and that
$\nabla_\mu$ transforms covariantly (adjoint representation) $\nabla_\mu
\rightarrow {\cal U}\nabla_\mu{\cal U}^{-1}$, one then sees $\hat{W}\rightarrow
{\cal U}\hat{W}{\cal U}^{-1}$.
\\
The Wigner equation is then an infinite-order differential equation. It is this 
equation which is the subject of this study.

\subsection*{Clifford Decomposition}
The Wigner operator takes values in the Clifford algebra, since it is given as a
product of two spinors. Hence it can be decomposed. In $d=4$ a basis for the
Clifford algebra is given by $1,\gamma_5,\gamma_a,\gamma_a\gamma_5,
\sigma_{ab}$, as is well-known. We thus write
\begin{equation}
    \hat{W} = {\cal S}+{\cal P}\gamma_5+{\cal V}^a\gamma_a +{\cal
    A}^a\gamma_a\gamma_5+{\cal T}^{ab}\sigma_{ab}
\end{equation}
where then, of course, $\cal S,P$ are a scalar and pseudo-scalar, ${\cal V}^a,
{\cal A}^a$, vector and axial-vector and ${\cal T}^{ab}$ an antisymmetric
tensor. A similar splitting-up of the operators can be made using the
properties of the Dirac matrices. The left hand side can be written
\begin{eqnarray}
    &&\left[m+\gamma^d\left(p_d+\frac{1}{2}ie_d^\mu\partial_\mu
    -\frac{1}{2}\omega_\mu^{bc}e^\mu_b\eta_{cd}\right)
    +2i\epsilon^a_{~bcd}e_a^\mu\omega_\mu^{bc}\gamma^5\gamma^d\right]\hat{W}
	\nonumber\\
    &&\equiv\left[m+\gamma^d A_d+\gamma_5\gamma^d B_d\right]\hat{W}
\end{eqnarray}
having used
\begin{displaymath}
	\gamma^d\sigma^{bc} = i(\eta^{ac}\eta^{bd}-\eta^{ab}\eta^{dc})\gamma_a
	+4\epsilon^{dcba}\gamma_5\gamma_a
\end{displaymath}
which follows from the definition of $\sigma_{ab}$ and the standard trace
formulas for the Dirac matrices, see e.g. Itzykson and Zuber \cite{IZ}.\\
We will write the right hand side as
\begin{equation}
	\hat{J}^{abc}\gamma_a\{\sigma_{bc},\hat{W}\}
\end{equation}
where $\hat{J}^{abc}$ then contains all the curvature information and no Clifford
algebra information (i.e. it is proportional to $1$).\\
With this notation we get
\begin{eqnarray*}
	m{\cal S}-A_a{\cal V}^a+B_a{\cal A}^a &=& -4\epsilon_{abcd}\hat{J}^{abc}
	{\cal A}^d\\
	m{\cal P}+A_a{\cal A}^a-B_a{\cal V}^a &=& -\eta_{ac}\eta_{bd} \hat{J}^{abc}
	{\cal A}^d\\
	m{\cal V}^d-A^d{\cal S}+B^d{\cal P}-i(\eta_{ef}\delta_g^d-\eta_{eg}
	\delta_f^d)A^e {\cal T}^{fg} &&\\
	\qquad +4\epsilon_{efg}^{~~~d}B^e{\cal T}^{fg} &=& i\eta_{ab}\delta_c^d
	\hat{J}^{abc}{\cal S}-2\epsilon_{abc}^{~~~d}\hat{J}^{abc}{\cal P}+\\
	&&4\delta_a^d \eta_{bf}\eta_{ce}\hat{J}^{abc}{\cal T}^{ef}\\
	m{\cal A}^d+A^d{\cal P}-B^d{\cal S}+i(\eta_{ef}\delta_g^d-\eta_{eg}
	\delta_f^d)B^e{\cal T}^{fg}&&\\
	\qquad -4\epsilon_{efg}^{~~~d}A^e{\cal T}^{fg} &=& -4\epsilon^d_{~abc}
	\hat{J}^{abc}{\cal S}+8i(\eta_{ab}\delta_c^d-\eta_{ac}\delta_b^d)
	\hat{J}^{abc}{\cal P}\\
	m{\cal T}^{ef}+\frac{1}{2}iA^{[e}{\cal V}^{f]}-\frac{1}{2}iB^{[e}
	{\cal A}^{f]} &=& 12\epsilon_{abc}^{~~~g}\epsilon^{ef}_{~~dg}
	\hat{J}^{abc}{\cal T}^{dg}-(\eta_{ac}\epsilon^{ef}_{~~db}-\\
	&&\qquad\eta_{da}
	\epsilon^{ef}_{~~cb})\hat{J}^{abc}{\cal A}^d
\end{eqnarray*}
where we have used the symmetry properties of $\hat{J}^{abc}$, i.e. $\hat{J}^{abc}
=-\hat{J}^{acb}$, to remove some terms involving $\eta$'s.\\
One will often be able to assume ${\cal P}={\cal A}^a=0$, in which case the 
equations can be simplified a little bit.\\
The equations involve not only the curvature 2-form but also its dual, since
\begin{equation}
	\epsilon_{abcd}\hat{J}^{abc} = -\frac{1}{2}i\kappa e^\mu_a(j_0-ij_1)
	\frac{\partial}{\partial p_\nu}(*R_{\mu\nu}^{ab})\eta_{bd}
\end{equation}
with
\begin{equation}
	*R_{\mu}^{ab} := \frac{1}{2}\epsilon_{abcd}R_{\mu\nu}^{cd}
\end{equation}
One knows that in Yang-Mills theory self-dual and anti self-dual curvature 2-forms
play an important role (they correspond to instantons, \cite{Ramond,GS,Nash,Nakahara}), it
is therefore encouraging that the dual of the curvature 2-form also appears in this
gravitational setting.\\
One should also note that the combination $\eta_{ab}\delta_c^d\hat{J}^{abc}$ is
proportional to the Ricci tensor, $\eta_{ab}\delta_c^d\hat{J}^{abc} = \frac{1}{2}
\pi^{-1/2}\kappa\frac{\partial}{\partial p_\nu}R_\nu^{~d}(j_0-ij_1)$,
which implies that this combination vanishes for
vacuum solution of general relativity, i.e. in Ricci-flat spacetimes $R_{\mu\nu} =
R_\mu^{~d}\eta_d^a e_\nu^a=0$. This
implies that in many backgrounds of real physical interest, the right hand sides
of these coupled equations simplify.\\
Another important consequence of this Clifford algebra decomposition concerns the
classical limit. In this limit one would expect the Wigner function, considered
as a $4\times 4$ matrix to be diagonal:
\begin{displaymath}
	\hat{W} = \left(\begin{array}{cccc} f_1 &&&\\ &f_2 &&\\ &&g_1 &\\ &&& g_2
	\end{array}\right) \equiv \left(\begin{array}{cc} F_+ & 0 \\ 0 & F_-
	\end{array}\right) \equiv {\cal S}~1+{\cal V}^0\gamma_0
\end{displaymath}
This leads to the following set of coupled equations
\begin{eqnarray}
	m{\cal S}-A_0{\cal V}^0 &=& 0\\
	B_0{\cal V}^0 &=& 0\\
	m{\cal V}^0-A^0{\cal S} &=& i\eta_{ab}\delta_c^0\hat{J}^{abc}{\cal S}\\
	B^d{\cal S} &=& 4\epsilon^d_{~abc}\hat{J}^{abc}{\cal S}
\end{eqnarray}
where, as we have seen, the right hand side in the third of these vanish if the
spacetime is Ricci-flat, i.e. a solution to the vacuum Einstein equations.\\
The second of these of equations imply ${\cal V}_0=0$, which then, from the first,
yields $m=0$, hence we are left with the following pair of equations
\begin{eqnarray}
	A_0{\cal S} &=& -i\eta_{ab}\delta_c^0\hat{J}^{abc}{\cal S}\\
	B^d{\cal S} &=& 4\epsilon^d_{~abc}\hat{J}^{abc}{\cal S}
\end{eqnarray}
Thus in a Ricci-flat spacetime, the only quantum corrections enter through the
second of these, as the first become $A_0{\cal S}=0$, i.e. the quantum
corrections will depend only on the dual of the curvature two-form. The first
equation implies, then, that $\cal S$ is an eigenstate of the differential
operator $e^\mu_0\partial_\mu+i\omega_\mu^{a0}e_\mu^a$ with eigenvalue $-2ip_0$,
whereas the second equation determines the dependency on the other components of
the momentum. We will return to the case $\hat{W} = {\cal S}$ in later sections.\\
Let us note in passing that the extra term $-2ip_0$ added to the Dirac operator
can be interpreted as a coupling to an Abelian connection, or as corresponding to
a Dirac operator in $5$ dimensions (as in the families index theorem, 
\cite{Nash,Nakahara}).\\
For later use, we also give the Clifford decomposition in $d=2$ dimensions. Here
$\gamma^a$ corresponds to $\sigma_1,\sigma_2$ (the Pauli matrices) and both 
$\gamma_5$ and $\sigma_{ab}$ corresponds to $\sigma_3 = -i\sigma_1\sigma_2 = 
-\frac{1}{2}i[\sigma_1,\sigma_2]$. The covariant derivative can be written as
$\nabla_\mu = \partial_\mu$, since there is no spinor connection in $d=2$, 
\cite{Nakahara}.
With this (Latin indices going from 1 to 2) the Wigner equation
reads
\begin{equation}
	\left[m-\sigma^i(p_i+\frac{1}{2}i\partial_i)
	\right]\hat{W} = -\frac{1}{4}i\kappa\sigma^i e_i^\alpha \hat{J}
	R_{\alpha\beta}\left\{\sigma_3,\frac{\partial}{\partial p_\beta}\hat{W}
	\right\}
\end{equation}
With the decomposition
\begin{equation}
	\hat{W} = W_0~1+W_i~\sigma^i+W_3~\sigma^3
\end{equation}
we then arrive at
\begin{eqnarray}
	mW_0-(p_i+\frac{1}{2}i\partial_i)W^i
	&=& -\frac{1}{2}i\kappa\epsilon^{ij}e_i^\alpha R_{\alpha\beta}\hat{J}
	\frac{\partial}{\partial p_\beta}W_j\\
	mW_i-\epsilon_{ij}W^j-(p_i+\frac{1}{2}i\partial_i)W_0 &=&
	-\frac{1}{2}i\kappa\left(e_i^\alpha R_{\alpha\beta}\hat{J}\frac{\partial}
	{\partial p_\beta}W_3+\epsilon_{ij}e^{j\alpha}R_{\alpha\beta}\hat{J}
	\frac{\partial}{\partial p_\beta}W_0\right)\nonumber\\
	&&\\
	mW_3-i\epsilon_{ij}(p^i+\frac{1}{2}i\partial^i)W^j
	&=& -\frac{1}{4}i\kappa\eta^{ij}e_i^\alpha R_{\alpha\beta}\hat{J}
	\frac{\partial}{\partial p_\beta}W_j
\end{eqnarray}
We will return to this set of equations in the section concerning the trace of the
energy momentum tensor in $d=2$ and $d=4$. For now let us just note that the
set of coupled equations in $d=2$ can be re-expressed as an equation for $W_i$ alone
by using the first and third to express $W_0, W_3$ as some operators acting on
$W_i$. Explicitly
\begin{eqnarray}
	W_0 &=& -\frac{i\kappa}{2m}\epsilon^{ij}e_i^\alpha R_{\alpha\beta}\hat{J}
	\frac{\partial W_j}{\partial p_\beta}+\frac{1}{m}(p_i+\frac{1}{2}i\partial_i)
	W^i\\
	W_3 &=& -\frac{i\kappa}{4m}\eta^{ij}e_i^\alpha R_{\alpha\beta}\hat{J}
	\frac{\partial W_j}{\partial p_\beta}+\frac{1}{m}\epsilon_{ij}(p^i+\frac{1}{2}
	i\partial^i)W^j
\end{eqnarray}
leading to
\begin{eqnarray}
	(m^2\eta_{ij}-m\epsilon_{ij})W^j-\frac{1}{m}(p_i+\frac{1}{2}i\partial_i)
	(p_j+\frac{1}{2}i\partial_j)W^j &\hspace{-2cm}& \nonumber\\
	+\frac{1}{2}i\kappa\epsilon^{kj}(p_i+\frac{1}{2}i\partial_i)\left(e_k^\alpha
	R_{\alpha\beta}\hat{J}\frac{\partial W^j}{\partial p_\beta}\right) &=&
	-\frac{1}{8}\kappa^2\eta^{kj}e_i^\alpha e_k^\gamma R_{\alpha\beta}R_{\gamma
	\delta}\hat{J}^2\frac{\partial^2 W^j}{\partial p_\beta\partial p_\delta}
	\nonumber\\
	&&-\frac{1}{4}\kappa^2\epsilon_i^j\epsilon^{kl}e_j^\alpha e_l^\gamma
	R_{\alpha\beta}R_{\gamma\delta}\hat{J}^2\frac{\partial^2 W_k}{\partial 
	p_\beta\partial p_\delta}\nonumber\\
	&&+\frac{1}{2}\kappa\epsilon_{kj} e_i^\alpha R_{\alpha\beta} (p^k+\frac{1}{2}
	i\partial^k)\hat{J}\frac{\partial W^j}{\partial p_\beta}\nonumber\\
	&&-\frac{1}{2}i\kappa\epsilon_i^je_j^\alpha R_{\alpha\beta} (p_k+\frac{1}{2}
	i\partial_k)\hat{J}\frac{\partial W^k}{\partial p_\beta}\nonumber\\
\end{eqnarray}
Remembering $\hat{J}=j_0(\frac{1}{2}\bigtriangleup)-ij_1(\frac{1}{2}\bigtriangleup)$
and that $\bigtriangleup$ is hermitian, the hermitian and anti-hermitian parts of
this equation must be satisfied separately, which then leads to two coupled 
equations which $W_i$ has to satisfy.

\subsection*{Example: Covariantly Constant Curvature}
When the curvature tensor is covariantly constant, $\nabla_\lambda R_{\mu\nu}^{ab} 
=0$, we can take the square of the equation for $\hat{W}$
\begin{displaymath}
	\left(m-\gamma_aA^a-\gamma_5\gamma_aB^a+\frac{1}{4}i\kappa e^\mu_a\gamma^a 
	R_{\mu\nu}^{bc}\left\{\sigma_{bc},\frac{\partial}{\partial p_\nu} \cdot
	\right\}\right)\hat{W}=0
\end{displaymath}
to obtain
\begin{eqnarray}
	0&=&\left[m^2-A^2-B^2-2\gamma_5 A\cdot B\right]\hat{W} +\nonumber\\
	&&\frac{\kappa^2}{16}e^\mu_ae^\rho_d\gamma^a R_{\mu\nu}^{bc}R_{\rho
	\sigma}^{ef}\frac{\partial^2}{\partial p_\nu\partial p_\sigma}
	\{\sigma_{bc},\gamma^d\{\sigma_{ef},\hat{W}\}\}-\nonumber\\
	&&\frac{1}{4}i\kappa\left\{\gamma_a A^a,e_d^\mu\gamma^d R_{\mu\nu}^{ef}
	\frac{\partial}{\partial p_\nu}\{\sigma_{ef},\hat{W}\}\right\}
	-\nonumber\\
	&&\frac{1}{4}i\kappa\left\{\gamma_5\gamma_aB^a,e^\mu_d\gamma^d
	R_{\mu\nu}^{ef}\frac{\partial}{\partial p_\nu}\{\sigma_{ef},\hat{W}\}
	\right\}
\end{eqnarray}
We can split this into two equations by taking the hermitian and anti-hermitian
parts out. For the hermitian part we then get the constraint equation generalising
the mass-shell condition
\begin{equation}
	0=(m^2-A^2-B^2-2A\cdot B\gamma_5)\hat{W} +\frac{\kappa^2}{16}
	e^\mu_ae^\rho_d\gamma^aR_{\mu\nu}^{bc}R_{\rho\sigma}^{ef}
	\frac{\partial^2}{\partial p_\nu\partial p_\sigma}
	\left\{\sigma_{bc},\gamma^d\left\{\sigma_{ef},\hat{W}\right\}\right\}
	\label{eq:diff}
\end{equation}
whereas the anti-hermitian part becomes the kinetic equation proper (as in the
Yang-Mills case, \cite{EGV})
\begin{equation}
	0=\left\{\gamma_aA^a+\gamma_5\gamma_aB^a,
	e^\mu_d\gamma^dR_{\mu\nu}^{ef}\frac{\partial}
	{\partial p_\nu}\{\sigma_{ef},\hat{W}\}\right\} \label{eq:kin}
\end{equation}
The mass-shell condition, (\ref{eq:diff}), can also be seen as a momentum diffusion
equation of the Fokker-Planck type \cite{kinetic,van}, but the important thing 
to note is that it
is {\em not} on its own a kinetic equation specifying the state of the matter 
fields. Comparing this with a standard classical kinetic equation for a phase space
distribution function $\cal F$, \cite{kinetic}
\begin{equation}
	\frac{\partial}{\partial p_\nu} (a_\nu{\cal F})+\frac{1}{2}
	\frac{\partial^2}{\partial p_\mu\partial p_\nu}(b_{\mu\nu}{\cal F})
	=f(p,x){\cal F}
\end{equation}
we see that $a_\nu$, the dynamical friction in our case is
\begin{equation}
	a_\nu = 0
\end{equation}
while $b_{\mu\nu}$, the momentum diffusion coefficient, is (simply take $\hat{W}={\cal
 S}$ to get the non-spin terms)
\begin{equation}
	b_{\nu\sigma} =  \frac{\kappa^2}{4}e^\mu_a e^\rho_d(\eta^{gf}\eta^{de}
	-\eta^{ge}\eta^{df})(\eta_{gc}\delta_b^a-\eta_{gb}\delta_c^a)
	R_{\mu\nu}^{bc}R_{\rho\sigma}^{ef}+\mbox{spin terms}
\end{equation}
and $f(p,x)$, a kind of momentum force term, finally reads
\begin{equation}
	f(p,x) = -(m^2-A^2-B^2)+\mbox{Clifford algebra terms}
\end{equation}
and is thus an operator (remember: $A\sim p+\partial+\omega$).\\
Thus in this particularly simple case, we see that the mass-shell condition can be
interpreted as a diffusion equation in momentum space. Now, this momentum diffusion is
provided by the quantum and curvature corrections to the mass, i.e. the difference
between $m^2$ and $p^2$, it is thus a pure quantum phenomenon. This is a kinetic
interpretation of the uncertainty principle and of the renormalisation of the mass.
It is intuitively pleasing to see that the momentum diffusion coefficient is given
simply by the curvature, and is thus a purely geometric object, even though 
its source is purely quantum.
For a curvature which is {\em not} covariantly constant, the mass-shell condition
contains higher derivatives with respect to the momentum, such terms do not have
a direct classical interpretation and therefore represent pure quantum effects.\\
The main difference between (\ref{eq:kin}) and the result for Yang-Mills theory
is the appearance of the anticommutator. As we have seen earlier, this 
anticommutator can be traced back to $\hat{W}$ being Clifford algebra-valued and
that the generators of the Lorentz algebra for Dirac fermions is precisely
$\sigma_{ab}$ which is an element of the Clifford algebra. Remember that the
elements $\gamma_a\gamma_b$ of any Clifford algebra $C(r,s)$ generate the Lie 
algebra $spin(r,s)$ which is isomorphic to $so(r,s)$. For $r=3,s=1$ this is the
Lorentz algebra in $d=3+1$ dimensions, while it for $r=s=1$ is the corresponding
algebra in $d=1+1$ dimensions, \cite{GS}. 
It is this interrelationship between the Clifford 
algebra and the Lorentz algebra (which is then the gauged algebra for gravitational
systems) which accounts for the anticommutator.\\
In the extreme, classical limit $\hat{W}={\cal S}$, the kinetic equation reduces to
\begin{eqnarray}
	0 &=& 2i(\eta_{af}\eta_{de}-\eta_{ae}\eta_{df})A^ae^{\mu d}R_{\mu\nu}^{ef}
	\frac{\partial}{\partial p_\nu} {\cal S}-\nonumber\\
	&&\epsilon_{ef}^{~~gh}(\eta_{ab}\eta_{dg}-\eta_{ag}\eta_{dh})B^a
	e^{\mu d}R_{\mu\nu}^{ef}\frac{\partial}{\partial p_\nu}{\cal S}
\end{eqnarray}
This can be seen as an equation involving only dynamical friction and not momentum
diffusion. Writing it on the classical form
\begin{equation}
	f_1(p,x){\cal F} = \frac{\partial}{\partial p_\nu}(a_\nu {\cal F})
\end{equation}
where the ``force term'', $f_1$, contains the operator $\bigtriangleup$, we see that
the dynamical friction is given by
\begin{eqnarray}
	a_\nu &=& (\eta_{af}\eta_{de}-\eta_{ae}\eta_{df})(p^a-\frac{1}{2}
	\omega_\rho^{ba})e^{\mu d}R_{\mu\nu}^{ef}-\nonumber\\
	&&\epsilon_{ef}^{~~gh}\epsilon^{p~~a}_{~qr}e^\rho_p e^{\mu d}\omega_\rho^{qr}
	R_{\mu\nu}^{ef}
\end{eqnarray}
It is very interesting to note that the resulting equations for $\hat{W}$ splits into
two set of equations, one containing only momentum diffusion and the other only 
dynamical friction. In both cases are the sources for the processes given by quantum
effects, i.e. vanish in the classical limit if one ignores spin effects. Furthermore,
the coefficients $a_\mu, b_{\mu\nu}$ are in both cases given by the curvature and are
thus geometrical quantities as one would expect intuitively.\\
Again, in the general case we would get higher derivatives of momentum, the 
``diffusion'' equation containing all even powers of $\frac{\partial}{\partial p}$, 
and the ``friction'' equation all the odd powers. These extra terms are pure quantum
effects and have no direct classical interpretation.\\
To round off the discussion we also give the splitting of the Wigner equation in 
$d=2$. As we have seen, in this case we can make do with an equation for $W_i$ alone.
The hermitian part of this equation is seen to be in the case of covariantly
constant curvature (corresponding to taking $\hat{J}\equiv 1$)
\begin{eqnarray}
	 (m^2\eta_{ij}-m\epsilon_{ij})W^j&&\nonumber\\
	-(p_ip_j+\frac{1}{2}i(p_i\partial_j
	+p_j\partial_i)-\frac{1}{4}\partial_i\partial_j)W^j
	&=& -\frac{1}{8}\kappa^2\eta^{kj}e_i^\alpha e_k^\gamma R_{\alpha\beta}
	R_{\gamma\delta}\frac{\partial^2 W^j}{\partial p_\beta\partial p_\delta}
	\nonumber\\
	&&-\frac{1}{4}\epsilon_i^j\epsilon^{kl}e_j^\alpha e_l^\gamma R_{\alpha\beta}
	R_{\gamma\delta}\frac{\partial^2 W_k}{\partial p_\beta \partial p_\delta}
	\nonumber\\
	&&+\frac{1}{2}\kappa\epsilon_{kj}e_i^\alpha R_{\alpha\beta}(p^k+\frac{1}{2}
	i\partial^k)\frac{\partial W^j}{\partial p_\beta}\nonumber\\
\end{eqnarray}
for the hermitian part, and
\begin{equation}
	\epsilon^{kj}(p_i+\frac{1}{2}i\partial_i)e_k^\alpha R_{\alpha\beta}
	\frac{\partial W_j}{\partial p_\beta} = \epsilon_i^j e_j^\alpha
	R_{\alpha\beta}(p^k+\frac{1}{2}i\partial^k)\frac{\partial W_j}{\partial 
	p_\beta}
\end{equation}
for the anti-hermitian part. Once again, we can interpret the hermitian part
as a momentum diffusion equation, both this time we also have a dynamical
friction contribution
\begin{displaymath}
	\frac{1}{2}\frac{\partial^2}{\partial p_\beta p_\delta}
	(b_{\beta\delta}^{kl}W_l)
	+\frac{\partial}{\partial p_\beta}(a_\beta^{kl}W_l)=
	f^{kl}(x,p)W_l
\end{displaymath} 
The diffusion coefficient is
\begin{equation}
	b_{\beta\delta}^{ik} = 
	\frac{1}{2}\kappa^2\left(\epsilon^{ij}\epsilon^{kl} e_j^\alpha e_l^\gamma
	+\frac{1}{2}\eta^{jk}\eta^{il}e_l^\alpha e_j^\gamma\right)R_{\alpha\beta}
	R_{\gamma\delta}
\end{equation}
while the dynamical friction becomes
\begin{equation}
	a_\beta^{kl} = -\frac{1}{2}\kappa\epsilon^{il}\eta^{jk}e_j^\alpha
	R_{\alpha\beta}(p_i+\frac{1}{2}i\partial_i)
\end{equation}
which is then a differential operator in this case. The force term is again related
to the mass-shell constraint, although in a somewhat more complicated form
\begin{equation}
	f^{kl} = m^2\eta^{kl}-m\epsilon^{kl}-p^kp^l-\frac{1}{2}i(p^k\partial^l
	-p^l\partial^k)+\frac{1}{4}\partial^k\partial^l
\end{equation}
The anti-hermit-an part is once more a pure dynamical friction equation (it would have
a momentum diffusion term too, if the curvature wasn't covariantly constant), but 
there is no source term. Thus there is quite a big difference between the results in
$d=2$ and $d=4$ dimensions, a difference which cannot simply be referred to the
different Lorentz character of the ``distribution'' in the two cases (in $d=2$ we 
considered a vector, and in $d=4$ a scalar), but is very much due to the difference
in the structure of the two Clifford algebras and the fact that there is no
spin connection in $d=2$. This is an important caveat.  	

\subsection*{The $\bigtriangleup$-Expansion: Quantum Corrections}
The case of covariantly constant curvature corresponds, as we have seen, to a
kind of classical limit involving only few quantum corrections. In this 
section we will commence a more systematic study of quantum corrections. This
is done by noting that $\bigtriangleup$ appears multiplied by $\hbar$, and that
Planck's constant only enters in this combination. An expansion in 
$\bigtriangleup$ is therefore related to an expansion in $\hbar$. Thus, one
can calculate quantum corrections by expanding the spherical Bessel functions
in $\hat{X}$ to a given order. To do this we need the standard formulae
\begin{eqnarray*}
	j_0(z) &=& \frac{\sin z}{z} = \sum_{n=0}^\infty (-1)^n\frac{z^{2n}}
	{(2n+1)!}\\
	j_1(z) &=& \frac{\sin z}{z^2}-\frac{\cos z}{z} = \sum_{n=1}^\infty
	(-1)^n\frac{2n}{(2n+1)!}z^{2n-1}
\end{eqnarray*}
Now, $\bigtriangleup$ is an hermitian operator and the Bessel functions only
appear in the combination $j_0-ij_1$, consequently it is rather simple to
make a separation into hermitian and anti-hermit-an parts of the Wigner
equation. One of these will then only contain odd powers of $\bigtriangleup$
and the other only even powers.\\ 
Due to the complicated nature of the operators appearing in the Wigner 
equation, we cannot simply take the square and compare with the classical
kinetic equations. Instead we have to make do with the equations on the
``Dirac form'' (in contrast to the ``Klein-Gordon form'' resulting from
taking squares), which do not have a direct classical analogue. The
interpretation will therefore not be as precise as one would perhaps have 
wanted.\\
Writing 
\begin{equation}
	W(x,p) = \sum_{n=0}^\infty \hbar^n W^{(n)}(x,p)
\end{equation}
an remembering $z=\frac{1}{2}\hbar\bigtriangleup$, we can collect terms with
the same power of Planck's constant. Doing this we get
\begin{equation}
	\left[m+e^\mu_a\gamma^a(p_\mu+\frac{1}{2}i\nabla_\mu)\right]
	W^{(0)} = -\frac{1}{2}i\kappa e^\mu_a\gamma^a R_{\mu\nu}^{bc}
	\left\{\frac{\partial}{\partial p_\nu}W^{(0)},\sigma^{bc}\right\}
\end{equation}
for the ``classical'' contribution, $\hbar^0$, 
which is recognised as the same as for the case of covariantly constant
curvature treated in the previous section, whereas the first correction
satisfies
\begin{eqnarray}
	\left[m+e^\mu_a\gamma^a(p_\mu+\frac{1}{2}i\nabla_\mu)\right]
	W^{(1)} &=& -\frac{1}{2}i\kappa e^\mu_a\gamma^a R_{\mu\nu}^{bc}
	\left\{\frac{\partial}{\partial p_\nu}W^{(1)},\sigma_{bc}\right\}
	+\nonumber\\
	&&\qquad\frac{1}{6}\kappa e^\mu_a\nabla_\rho(\gamma^aR_{\mu\nu}^{bc})
	\left\{\frac{\partial^2}{\partial p_\nu\partial p_\rho}W^{(0)},
	\sigma_{bc}\right\}\nonumber\\
\end{eqnarray}
while the second order contribution satisfies (remembering that the vierbein
is covariantly constant)
\begin{eqnarray}
	\left[m+e^\mu_a\gamma^a(p_\mu+\frac{1}{2}i\nabla_\mu)\right]
	W^{(2)} &=& -\frac{1}{2}i\kappa e^\mu_a\gamma^a R_{\mu\nu}^{bc}
	\left\{\frac{\partial}{\partial p_\nu}W^{(2)},\sigma_{bc}\right\}
	+\nonumber\\
	&&\frac{1}{6}\kappa e^\mu_a\nabla_\rho(\gamma^aR_{\mu\nu}^{bc})
	\left\{\frac{\partial^2}{\partial p_\nu\partial p_\rho}W^{(1)},
	\sigma_{bc}\right\}+\nonumber\\
	&&\frac{1}{48}i\kappa e^\mu_a\nabla_\rho\nabla_\sigma(\gamma^a 
	R_{\mu\nu}^{bc})\left\{\frac{\partial^3}{\partial p_\nu\partial p_\rho
	\partial p_\sigma}W^{(0)},\sigma_{bc}\right\}\nonumber\\
\end{eqnarray}
In general we will have a recursive scheme
\begin{eqnarray}
	&&\left[m+e^\mu_a\gamma^a(p_\mu+\frac{1}{2}i\nabla_\mu)\right]
	W^{(n)} +\frac{1}{2}i\kappa e^\mu_a\gamma^aR_{\mu\nu}^{bc}
	\left\{\frac{\partial}{\partial p_\nu}W^{(n)},\sigma_{bc}\right\}
	\nonumber\\
	&&\qquad = \mbox{terms involving only }W^{(k)}\mbox{ with }k<n
\end{eqnarray}
The terms on the right hand side will involve more and more momentum
derivatives of the Wigner functions $W^{(k)}$ and similarly higher and higher
order covariant derivatives of the vierbeins and curvature two-forms. Higher
order derivatives of curvature can in general be considered as related to
fluctuations of the geometry, thus the higher order Wigner functions are
determined by the fluctuations of spacetime, moreover they couple to
higher order derivatives with respect to the momentum of the lower order
Wigner functions, terms which do therefore not have a classical interpretation
(only first and second order momentum derivatives appear in classical kinetic
equations). This shows how the expansion in $\bigtriangleup$ is closely
related to pure quantum effects with no classical analogue. One cannot, 
however, guarantee that this expansion is equivalent to the standard loop
expansion in quantum field theory. In general there will be no such simple
relationship.\\
One should also take notice of the fact that the momentum derivative operator
is symmetric in its indices, this implies that only the symmetric part
of $\nabla_\rho...\nabla_\sigma$ acting on the curvature will contribute.
From the commutator relation we get
\begin{equation}
	\nabla_{(\mu}\nabla_{\nu)} = \nabla_\mu\nabla_\nu-R_{\mu\nu}^{ab}
	\sigma_{ab}
\end{equation}
whence it follows that the right hand side of the recursive scheme not only
includes higher and higher order derivatives of the curvature but also
higher and higher powers of it.\\
If we introduce the operator $\hat{Y}$ by
\begin{equation}
	\hat{Y}W := \left[m+e^\mu_a\gamma^a(p_\mu+\frac{1}{2}i\nabla_\mu)
	+\frac{1}{2}i\kappa e^\mu_a\gamma^aR_{\mu\nu}^{bc}\left\{
	\frac{\partial}{\partial p_\nu}\cdot,\sigma_{bc}\right\}\right]W
\end{equation}
then we can write the recursive scheme as
\begin{equation}
	\hat{Y}W^{(n)} = F^{(n)}
\end{equation}
where $F^{(0)}=0$ and $F^{(n)}$ depends on $W^{(k)},k<n$. Thus, on the formal
level
\begin{equation}
	W^{(n)}(x,p) = \hat{Y}^{-1}F^{(n)}(x,p)\qquad ,\qquad n\geq 1
\end{equation}
Written out explicitly, the first corrections then read
\begin{eqnarray}
	W^{(1)} &=& \frac{1}{6}\kappa\hat{Y}^{-1} e^\mu_a\nabla_\rho(\gamma^a
	R_{\mu\nu}^{bc})\left\{\frac{\partial^2W^{(0)}}{\partial p_\nu
	\partial p_\rho},\sigma_{bc}\right\}\\
	W^{(2)} &=& \frac{1}{6}\kappa\hat{Y}^{-1} e^\mu_a\nabla_\rho(\gamma^a
	R_{\mu\nu}^{bc})\left\{\frac{\partial^2W^{(1)}}{\partial p_\nu
	\partial p_\rho},\sigma_{bc}\right\}+\nonumber\\
	&&\frac{i\kappa}{48}\hat{Y}^{-1}e^\mu_a\nabla_\rho\nabla_\sigma
	(\gamma^aR_{\mu\nu}^{bc})\left\{\frac{\partial^3W^{(0)}}{\partial 
	p_\nu\partial p_\rho\partial p_\sigma},\sigma_{bc}\right\}
\end{eqnarray}
Defining
\begin{equation}
	H_{\rho\nu}^{bc}:=e^\mu_a\nabla_\rho(\gamma^aR_{\mu\nu}^{bc})
\end{equation}
we can rewrite this as
\begin{eqnarray}
	W^{(1)} &=& \frac{1}{6}\kappa \hat{Y}^{-1}H_{\rho\nu}^{bc}
	\left\{\frac{\partial^2W^{(0)}}{\partial p_\rho\partial p_\nu},
	\sigma_{bc}\right\}\\
	W^{(2)} &=& (\frac{1}{6}\kappa)^2\hat{Y}^{-1}H_{\rho\nu}^{bc}
	\left\{\frac{\partial^2}{\partial p_\rho\partial p_\nu}\hat{Y}^{-1}
	H_{\kappa\epsilon}^{gh}\left\{\frac{\partial^2W^{(0)}}{\partial
	p_\kappa\partial p_\epsilon},\sigma_{gh}\right\},\sigma_{bc}\right\}
	\nonumber\\
	&&+\frac{i\kappa}{48}\hat{Y}^{-1}\nabla_\rho H_{\sigma\nu}^{bc}
	\left\{\frac{\partial^3W^{(0)}}{\partial p_\nu\partial p_\rho
	\partial p_\sigma},\sigma_{bc}\right\}
\end{eqnarray}
And so on.\\
Let us Clifford decompose $\hat{Y}W$. This is straightforward and we get
\begin{displaymath}
	m{\cal S}-A^a{\cal V}_a+B^a{\cal A}_a+2i\kappa\epsilon^a_{~bcd}e^\mu_a
	R_{\mu\nu}^{bc}\frac{\partial}{\partial p_\nu}{\cal A}^d
\end{displaymath}
for the scalar part,
\begin{displaymath}
	m{\cal P}+A^a{\cal A}_a-B^a{\cal V}_a+\frac{1}{2}i\kappa \delta^a_c
	\eta_{bd}e^\mu_a
	R_{\mu\nu}^{bc}\frac{\partial}{\partial p_\nu}{\cal A}^d
\end{displaymath}
for the pseudo-scalar contribution,
\begin{eqnarray*}
	&&m{\cal V}^d-A^d{\cal S}+B^d{\cal P}-i(\eta_{ef}\delta^d_g-\eta_{eg}
	\delta^d_f)A^e{\cal T}^{fg}+\\
	&&4\epsilon_{efg}^{~~~~d}B^e{\cal T}^{fg}-\frac{1}{2}i\kappa e^\mu_a
	R_{\mu\nu}^{bc}\frac{\partial}{\partial p_\nu}\left[ -i\delta_b^a\delta^d_c
	{\cal S}-
	2\epsilon^{a~~~d}_{~bc}{\cal P}+4\eta^{ad}{\cal T}_{bc}\right]
\end{eqnarray*}
for the vector part, while the axial vector contribution turns out to be
\begin{eqnarray*}
	&&m{\cal A}^d+A^d{\cal P}-B^d{\cal S}+i(\eta_{ef}\delta^d_g-\eta_{eg}
	\delta^d_f)B^e{\cal T}^{fg}-\\
	&&4\epsilon_{efg}^{~~~~d}A^e{\cal T}^{fg}-\frac{1}{2}i\kappa e^\mu_a
	R_{\mu\nu}^{bc}\frac{\partial}{\partial p_\nu}
	\left[-4\epsilon^{da}_{~~bc}{\cal S}+8i(\delta^a_b\delta^d_c-\delta^a_c
	\delta^d_b){\cal P}\right]
\end{eqnarray*}
and finally
\begin{eqnarray*}
	&& m{\cal T}^{ef}+\frac{1}{2}iA^{[e}{\cal V}^{f]} -\frac{1}{2}iB^{[e}
	{\cal A}^{f]}-\\
	&&\frac{1}{2}i\kappa e^\mu_aR_{\mu\nu}^{bc}\frac{\partial}{\partial p_\nu}
	\left[12\epsilon^{a~~~g}_{~bc}\epsilon^{ef}_{~~dg}{\cal T}^{dg}
	+(\delta^a_c\epsilon^{ef}_{~~db}-\delta^a_d\epsilon^{ef}_{~~cb})
	{\cal A}^d\right]
\end{eqnarray*}
for the tensor part. On the right hand side of the recursion relation we have terms
of the form
\begin{displaymath}
	\frac{1}{6}\kappa e^\mu_a\nabla_\rho(\gamma^aR_{\mu\nu}^{bc})
	\left\{\frac{\partial^2}{\partial p_\nu\partial p_\rho}W,\sigma_{bc}
	\right\}
\end{displaymath}
Now, here the covariant derivative is to act on the two form $R_{\mu\nu}^{bc}$, 
i.e.
\begin{displaymath}
	\nabla_\rho R_{\mu\nu}^{bc} = \partial_\rho R_{\mu\nu}^{bc}+\Gamma^\lambda
	_{\rho\mu}R_{\lambda\nu}^{bc}+\Gamma^\lambda_{\rho\nu}R_{\lambda\mu}^{bc}
\end{displaymath}
and hence $\nabla_\rho$ doesn't involve any Clifford algebra elements 
($\sigma_{bc}$ is only the generator of $so(3,1)$ in the spin $\frac{1}{2}$ 
representation). Therefore we can move the (constant) Dirac matrix $\gamma^a$
outside the covariant derivation. A Clifford decomposition of this term is
thus straightforward, and we obtain
\begin{eqnarray*}
	\mbox{S}&:& -\frac{2}{3}\kappa e^\mu_a\epsilon^a_{~bcd}(\nabla_\rho
	R_{\mu\nu}^{bc})\frac{\partial^2{\cal A}^d}{\partial p_\nu\partial 
	p_\rho}\\
	\mbox{P}&:& -\frac{1}{6}\kappa e^\mu_a\eta_{bd}\delta^a_c(\nabla_\rho
	R_{\mu\nu}^{bc})\frac{\partial^2{\cal A}^d}{\partial p_\nu\partial 
	p_\rho}\\
	\mbox{V}&:& \frac{1}{6}i\kappa\delta^a_b\delta^d_ce^\mu_a(\nabla_\rho
	R_{\mu\nu}^{bc})\frac{\partial^2{\cal S}}{\partial p_\nu\partial p_\rho}
	-\frac{1}{3}\kappa\epsilon^{a~~~d}_{~bc}e^\mu_a(\nabla_\rho 
	R_{\mu\nu}^{bc})\frac{\partial^2{\cal P}}{\partial p_\nu\partial p_\rho}+\\
	&&\qquad \frac{2}{3}\kappa\eta^{ad}\eta_{bf}\eta_{ce}e^\mu_a(\nabla_\rho
	R_{\mu\nu}^{bc})\frac{\partial^2}{\partial p_\nu\partial p_\rho}
	{\cal T}^{ef}\\
	\mbox{A}&:& -\frac{2}{3}\epsilon^{da}_{~~bc}e^\mu_a(\nabla_\rho 
	R_{\mu\nu}^{bc})\frac{\partial^2{\cal S}}{\partial p_\nu\partial p_\rho}+\\
	&&\qquad \frac{4}{3}i\kappa(\delta^a_v\delta^d_c-\delta^a_c\delta^d_b)
	(\nabla_\rho R_{\mu\nu}^{bc})\frac{\partial^2{\cal P}}{\partial p_\nu	
	\partial p_\rho}\\
	\mbox{T}&:& 2\kappa e^\mu_a\epsilon^{a~~~g}_{~bc}\epsilon^{ef}_{~~dg}
	(\nabla_\rho R_{\mu\nu}^{bc})\frac{\partial^2}{\partial p_\nu\partial p_\rho}
	{\cal T}^{dg}-\\
	&&\qquad \frac{1}{6}\kappa (\delta^a_c\epsilon^{ef}_{~~db}-\delta^a_d
	\epsilon^{ef}_{~~cb})(\nabla_\rho R_{\mu\nu}^{bc})
	\frac{\partial^2{\cal A}^d}{\partial p_\nu\partial p_\rho}
\end{eqnarray*}
To have a look at a solution we can take the extreme case
\begin{equation}
	W^{(0)}(x,p) = {\cal S}_0(x,p) = {\cal N} e^{-\alpha^{\mu\nu}(x)p_\mu p_\nu
	+\beta^\mu(x)p_\mu+\gamma(x)}
\end{equation}
As we have seen earlier, this is only possible provided $m=0$, in which case
${\cal S}_0$ has to satisfy the coupled set of equations
\begin{eqnarray}
	-A^d{\cal S}_0+\frac{1}{2}\kappa e^\mu_a R_{\mu\nu}^{ad}
	\frac{\partial}{\partial p_\nu}{\cal S}_0&=&0\\
	-B^d{\cal S}_0+2i\kappa e^\mu_a R_{\mu\nu}^{bc}\epsilon^{da}_{~~bc}
	\frac{\partial}{\partial p_\nu}{\cal S}_0&=&0
\end{eqnarray}
The second of these yields upon insertion of the explicit form for ${\cal S}_0$
\begin{equation}
	\omega_\mu^{bc} = -\kappa R_{\mu\nu}^{bc}(\alpha^{\nu\rho}p_\rho+\beta^\nu)
\end{equation}
which implies $R_{\mu\nu}^{ab}\alpha^{\nu\rho}\equiv 0$.
Inserting this into the first we arrive at
\begin{equation}
	p_d+\frac{1}{2}ie^\mu_d(-\partial_\mu\alpha^{\nu\rho} p_\nu p_\rho+
	\partial_\mu\beta^\nu p_\nu+\partial_\mu\gamma)=0
\end{equation}
which is only possible if
\begin{eqnarray}
	\alpha^{\mu\nu} &=& 0\\
	\beta^\nu &=& 2i x^\nu+const.\\
	\gamma &=& const.
\end{eqnarray}
Thus ${\cal S}_0$ is of the so-called J\"{u}ttner form \cite{van}
\begin{displaymath}
	\exp(\beta(x)(\mu(x)-p^\mu U_\mu(x))
\end{displaymath}
where $\beta$ is the inverse temperature, $\mu$ the Gibbs energy and $U_\mu$ the
four velocity, i.e. $\beta^\mu = \beta U^\mu, ~~\gamma=\beta\mu$.
Let us now make a similar {\em Ansatz} for the first quantum correction, i.e.
$W^{(1)} \equiv {\cal S}_1 = {\cal N}' \exp(-\bar{\alpha}^{\mu\nu} p_\mu p_\nu
+\bar{\beta}^\mu p_\mu+\bar{\gamma})$. The equations then reduce to
\begin{eqnarray}
	-A^d{\cal S}_1+\frac{1}{2}\kappa e^\mu_a R_{\mu\nu}^{ad} \frac{\partial}
	{\partial p_\nu}{\cal S}_1&=& \frac{1}{6}i\kappa e^\mu_a(\nabla_\rho
	R_{\mu\nu}^{ad})\frac{\partial^2}{\partial p_\nu\partial p_\rho}{\cal S}_0\\
	-B^d{\cal S}_1+2i\kappa e^\mu_a R_{\mu\nu}^{bc}\epsilon^{da}_{~~bc}
	\frac{\partial}{\partial p_\nu}{\cal S}_1&=& -\frac{2}{3}\kappa
	\epsilon^{da}_{~~bc}e^\mu_a(\nabla_\rho R_{\mu\nu}^{bc})
	\frac{\partial^2}{\partial p_\nu\partial p_\rho}{\cal S}_0
\end{eqnarray}
then evaluating the differentiations with respect to the momenta and collecting powers
of these, we arrive at the following set of conditions
\begin{eqnarray}
	\partial_\mu\bar{\alpha}^{\nu\rho} &=& 0\\
	p_d-\frac{1}{2}ie^\mu_d(\partial_\mu\bar{\beta}^\nu) p_\nu
	+\frac{1}{2}\kappa e^\mu_a R_{\mu\nu}^{ad}\bar{\alpha}^{\nu\rho} p_\rho&=&0\\
	e_d^\mu\partial_\mu\bar{\gamma} &=& 0\\
	\frac{1}{2}e^\mu_a\left(\omega_\mu^{bc}\delta^a_b\eta_{cd}+\kappa 
	R_{\mu\nu}^{ad}\right){\cal S}_1&=& \frac{1}{6}i\kappa e^\mu_a(\nabla_\rho
	R_{\mu\nu}^{ad})\alpha^{\rho\nu}{\cal S}_0
\end{eqnarray}
from the first of the equations for ${\cal S}_1$, while the second give us
\begin{equation}
	(-i\epsilon^a_{~bcd}e^\mu_a\omega_\mu^{bc}+i\kappa e^\mu_a R^{bc}_{\mu\nu}
	\epsilon^{da}_{~~bc}(\bar{\beta}^\nu-\bar{\alpha}^{\nu\rho}p_\rho){\cal S}_1
	= \frac{1}{3}\kappa\epsilon^{da}_{~~bc}e^\mu_a(\nabla_\rho R_{\mu\nu}^{bc})
	\alpha^{\nu\rho}{\cal S}_0
\end{equation}
Combining these we get the following set of conditions
\begin{eqnarray*}
	\partial_\mu\bar{\alpha}^{\nu\rho} &=& 0\\
	R_{\mu\nu}^{bc}\bar{\alpha}^{\nu\rho} &=& 0\\
	ie^\mu_a(\partial_\mu\bar{\beta}^\nu)p_\nu-\kappa e^\mu_aR_{\mu\nu}^{ac}
	\eta_{cd}\bar{\alpha}^{\nu\rho}p_\rho &=& 2p_d\\
	(\epsilon^a_{~bcd}e^\mu_a\omega_\mu^{bc}-\kappa e^\mu_aR_{\mu\nu}^{bc}
	\epsilon^{da}_{~~bc}\bar{\beta}^\nu){\cal S}_1 &=& \frac{1}{3}\kappa
	\epsilon^{da}_{~~bc}e^\mu_a(\nabla_\rho R_{\mu\nu}^{bc})\alpha^{\nu\rho}
	{\cal S}_0
\end{eqnarray*}
In a general spacetime the first two of these will give $\bar{\alpha}^{\mu\nu}\equiv 
0$ as for the lowest order term. Hence the solution ${\cal S}_1$ will once
more be of the J\"{u}ttner form, albeit with a much more complicated expression
for $\beta(x)U^\mu(x) = \bar{\beta}^\mu$.\\ 
To get more information about the recursive scheme presented in this section, we can
find the hermitian and anti-hermitian parts of the squared equations (i.e. the 
Wigner equation on ``Klein-Gordon form''), as this is where we expect to see the
analogy with classical kinetic theory most clearly.\\
We essentially calculated the square of $\hat{Y}$ in the previous section, the
square of $F^{(1)}$ is similarly
\begin{equation}
	(F^{(1)})^2=
	\frac{\kappa^2}{36}H_{\nu\rho}^{bc}\left\{\frac{\partial^2 W^{(0)}}
	{\partial p_\nu\partial p_\rho},\sigma_{bc}\right\}
	H_{\kappa\epsilon}^{gh}\left\{\frac{\partial^2 W^{(0)}}
	{\partial p_\kappa\partial p_\epsilon},\sigma_{gh}\right\}
\end{equation}
A Clifford decomposition of this is complicated by the $\gamma^a$ part of $H^{bc}
_{\mu\nu}$ and the quadratic appearance of $W^{(0)}$. We note that $(F^{(1)})^2$ is
hermitian, hence it only contributes to the momentum diffusion equation generalising
the mass-shell constraint, whereas the kinetic equation proper (which comes from the
anti-hermitian part of $\hat{Y}^2$) does not get any contribution from lower order
terms. The momentum diffusion equation then reads
\begin{eqnarray}
	(m^2-A^2-B^2-2A\cdot B\gamma_5)W^{(1)} &&\nonumber\\
	&&+\frac{\kappa^2}{16}
	e^\mu_ae^\rho_d\gamma^aR_{\mu\nu}^{bc}R_{\rho\sigma}^{ef}
	\frac{\partial^2}{\partial p_\nu\partial p_\sigma}
	\left\{\sigma_{bc},\gamma^d\left\{\sigma_{ef},W^{(1)}\right\}\right\}
	\nonumber\\
	&=& \frac{\kappa^2}{36}H_{\nu\rho}^{bc}\left\{\frac{\partial^2 W^{(0)}}
	{\partial p_\nu\partial p_\rho},\sigma_{bc}\right\}
	H_{\kappa\epsilon}^{gh}\left\{\frac{\partial^2 W^{(0)}}
	{\partial p_\kappa\partial p_\epsilon},\sigma_{gh}\right\}\nonumber\\
\end{eqnarray}
At the next order, $O(\hbar^2)$, however, there will be contributions from $W^{(0)},
W^{(1)}$ to the kinetic equation proper for $W^{(2)}$ too, namely
\begin{displaymath}
	\frac{1}{288}\kappa^2 
	\left\{H_{\rho\nu}^{bc}\left\{
	\frac{\partial^2 W^{(1)}}{\partial p_\nu\partial p_\rho}
	,\sigma_{bc}\right\},(\nabla_\sigma H_{\kappa\epsilon}^{ef})
	\left\{\frac{\partial^3 W^{(0)}}{\partial p_\sigma
	\partial p_\kappa\partial p_\epsilon},\sigma_{ef}\right\}\right\}
\end{displaymath}
In general, the momentum diffusion equation will contain only squares, whereas the kinetic
equation will contain only cross products (always in the form of an anticommutator)
of the lower order Wigner functions.

\section*{The Conformal Anomaly in $d=2$ and $d=4$}
Let us calculate the trace of $\langle T_{\mu\nu}\rangle$, 
denoted by $\langle T\rangle$. The non-vanishing of this quantity is
the conformal anomaly when $m=0$, \cite{QFT}.\\
The general expression for $\langle T_{ab}\rangle$ in terms of the Wigner
function is given by
\begin{equation}
	\langle T_{ab}(x)\rangle = {\rm Tr}~\int_{T_x^*M}
	\gamma_ap_b\langle \hat{W}\rangle d^dp
\end{equation}
It follows that (in $d$ dimensions, $W=\langle\hat{W}\rangle$)
\begin{equation}
	\langle T\rangle := \langle T_{ab}\eta^{ab}\rangle = {\rm Tr}~\int_{T_x^*M}
	\gamma^ap_aW d^dp
\end{equation}
Now, from the equation for $W$
\begin{displaymath}
	\left[m-\frac{1}{2}i\gamma^a(2ip_a+e_a^\mu\nabla_\mu)\right]W = \hat{X} W
\end{displaymath}
it follows that
\begin{equation}
	\gamma^ap_a W = (\frac{1}{2}i\gamma^aa_a^\mu\nabla_\mu-m)W-\hat{X}W
\end{equation}
Hence
\begin{equation}
	\langle T\rangle = {\rm Tr}~\int_{T_x^*M}\left[(\frac{1}{2}i\gamma^ae_a^\mu
	\nabla_\mu-m)W-\hat{X}W \right]d^dp
\end{equation}
We also know
\begin{equation}
	\hat{X}W = \frac{\partial}{\partial p_\nu}(W\times(\mbox{other terms}))
\end{equation}
i.e. that $\hat{X}W$ is a total derivative, and thus that its integral over the
cotangent space at $x$ vanish, leaving us with
\begin{equation}
	\langle T\rangle = {\rm Tr}~\int_{T_x^*M}(\frac{1}{2}i\gamma^ae_a^\mu
	\nabla_\mu-m)Wd^dp
\end{equation}
From this we see that $\langle T\rangle$ measures the failure of $W$ to satisfy a
Dirac equation with mass $2m$ (thus it is most interesting when $m=0$).
This equation is valid for all $d$. To carry out the trace we need the Clifford
decomposition, which is then $d$ dependent.\\
For $d=2$ we get
\begin{equation}
	\langle T(x)\rangle = \frac{1}{2}i\int_{T_x^*M} e^\mu_a\eta^{ab}
	\partial_\mu W_b d^2p-m\int_{T_x^*M} W_0d^2p
\end{equation}
where we have written $W=W_0~1+W_a~\sigma^a+W_3~\sigma_3$ with $a,b=1,2$.
Introducing the current density $\langle j_a(x)\rangle :=\int W_a d^2p$ and the number
density $\langle n(x)\rangle :=\int W_0d^2p$ we can write this in terms of 
these macroscopic quantities only as
\begin{equation}
	\langle T(x)\rangle = \frac{1}{2}ie_a^\mu\eta^{ab}\langle \partial_\mu j_b(x)
	\rangle -m~\langle n(x)\rangle
\end{equation}
For massless fields, the conformal anomaly is then related to the non-conservation
of the current $j_a$, such that $\langle T\rangle \neq 0$ if and only if $j_a$ is
not conserved in expectation value.\\
In $d=4$ we similarly get
\begin{eqnarray}
	\langle T(x)\rangle &=& \int_{T_x^*M} \left[\frac{1}{2}(ie^\mu_d\partial_\mu-
	\omega_\mu^{bc}e^\mu_b\eta_{cd}){\cal V}^d+2i\epsilon^a_{~bcd}e^\mu_a
	\omega_\mu^{bc}{\cal A}^d-m{\cal S}\right] d^4p\nonumber\\
	& :=& \frac{1}{2}(ie^\mu_d
	\partial_\mu-\omega_\mu^{bc}e^\mu_b\eta_{cd})\langle j^d\rangle +
	\epsilon^a_{~bcd}e^\mu_a\omega_\mu^{bc} \langle k^d\rangle -m
	\langle n\rangle
\end{eqnarray}
where we have introduced the vector current $\langle j_a\rangle 
:= \int {\cal V}_a d^4p$, the axial current $\langle k_a \rangle
:= \int {\cal A}_ad^4p$ and the number density $\langle n \rangle:=
\int {\cal S}d^4p$. We see that in $d\neq 2$ the possible existence of a 
conformal anomaly
or not is not as simply related to the question of the 
conservation of a current as in $d=2$.\\
The fact that an anomaly can be expressed in terms of the Wigner function 
integrated over the cotangent space, suggests a closer relationship between
$W$ and anomalies, in particular with the spin complex, \cite{Nash,Nakahara}. The
very nature of the Wigner function, or rather the entire Wigner-Weyl-Moyal
formalism where operators are replaced by symbols on the cotangent bundle,
makes the translation of analytical properties into geometrical or algebraic 
ones
possible. Such a translation is at the heart of index theorems, and it has in
fact been shown that the Atiyah-Singer index theorem can be related to the
classical limit of the Wigner-Weyl-Moyal (\small{WWM}) formalism for the 
ordinary 
Heisenberg algebra, \cite{RNN}. The way in which this generalised \small{WWM}
formalism relates to index theorems is presently under study.

\section*{The Hydrodynamic Equations: Moments}
The previous section calculated the trace of the energy-momentum tensor from the
Wigner equation, this is only one macroscopic quantity which one can define. Let
us define the following moments of the Wigner function (for simplicity we suppress
the $\langle\cdot\rangle$)
\begin{eqnarray*}
	n(x) &:=& {\rm Tr}~\int W d^4p = \int {\cal S} d^4p\\
	U^a(x) &:=& {\rm Tr}~\int p^a Wd^4p = \int p^a {\cal S}d^4p\\
	j_a(x) &:=& {\rm Tr}~\int\gamma_a Wd^4p = \int {\cal V}_a d^4p\\
	k_a(x) &:=& {\rm Tr}~\int\gamma_5\gamma_a Wd^4p = \int {\cal A}_a d^4p\\
	T_{ab}(x) &:=& {\rm Tr}~\int p_a\gamma_b Wd^4p = \int p_a{\cal V}_b d^4p\\
	s_{ab}(x) &:=& {\rm Tr}~\int \sigma_{ab}Wd^4p = \int {\cal T}_{ab}d^4p\\
	\tau_{ab}(x) &:=& {\rm Tr}~\int p_a\gamma_5\gamma_b Wd^4p = \int p_a
		{\cal A}_bd^4p\\
	\chi(x) &:=& {\rm Tr}~\int \gamma_5Wd^4p = \int {\cal P}d^4p\\
	\chi_a(x) &:=& {\rm Tr}~\int p_a\gamma_5Wd^4p = \int p_a{\cal P}d^4p\\
	\lambda_{abc}(x) &:=& {\rm Tr}~\int p_a\sigma_{bc}Wd^4p = \int p_a{\cal 
		T}_{bc}d^4p
\end{eqnarray*}
These are all the zeroth and first moments of the Wigner function in $d=4$ 
dimensions. Most of these quantities have a direct physical interpretation, $n(x)$
is the number or energy density, $U^a$ is a momentum density, $j_a$ a current,
$k_a$ and axial current, $T_{ab}$ the energy-momentum tensor and  $s_{ab}$ the spin
density, while $\tau_{ab}$ is a kind of ``pseudo-energy-momentum tensor'', $\chi$ 
and $\chi_a$ are related to chirality and $\lambda_{abc}$ represents a spin-momentum
interaction term. The equations of motion for these macroscopic quantities constitute
the corresponding set of (quantum) hydrodynamic equations, which are derived by
simply taken zeroth and first moments of the Wigner equation in its Clifford
decomposed form. Hence each of the five equations in this Clifford decomposition
gives rise to two equations for these moments. If we write
\begin{equation}
	A_a = p_a+\frac{1}{2}i\hat{D}_a \qquad J_a^{~bc} = \hat{J} e^\mu_a 
	R_{\mu\nu}^{bc} \frac{\partial}{\partial p_\nu}
\end{equation}
we get the following ten equations
\begin{eqnarray}
	mn-T-\frac{1}{2}i\hat{D}_aj^a+B_ak^a &=& 0 \label{eq:confan}\\
	mU_b-\int p_bp_a{\cal V}^ad^4p+\frac{1}{2}i\hat{D}_a T^{~a}_b
	+B_a\tau^{~a}_b &=& \epsilon^a_{~b'cd}\hat{J} e^\mu_aR_{\mu\nu}^{b'c}e^\nu_b
	k^d\\
	m\chi+\tau+\frac{1}{2}i\hat{D}_ak^a-B_aj^a &=& 0 \label{eq:psconfan}\\
	m\chi_b+\int p_ap_b{\cal A}^ad^4p+\frac{1}{2}i\hat{D}_a\tau^{~a}_b
	-B_aT^{~a}_b &=& \eta_{b'd}\hat{J}e^\mu_aR_{\mu\nu}^{b'c}e^\nu_b k^d\\
	mj^d-U^d-\frac{1}{2}i\hat{D}^dn+B^d\chi \qquad&&\nonumber\\
	-(\eta_{ef}\delta^d_g-\eta_{eg}\delta^d_f)(\lambda^{efg}+\frac{1}{2}i\hat{D}^e
	s^{fg})+4\epsilon_{efg}^{~~~d}B^es^{fg} &=& 0 \label{eq:heat}\\
	mT_h^{~d}-\int p_hp^d{\cal S}d^4p -\frac{1}{2}i\hat{D}^dU_h+B^d\chi_h
		\qquad &&\nonumber\\
	-i(\eta_{ef}\delta^d_g-\eta_{eg}\delta^d_f)(\int p_hp^e{\cal T}^{fg}d^4p
	+\frac{1}{2}i\hat{D}^e\lambda_h^{~fg})&&\nonumber\\
	+4\epsilon_{efg}^{~~~d}B^e\lambda_h^{~fg} &=& -i\delta^d_c\hat{J}e^\mu_b
	R_{\mu\nu}^{bc}e^\nu_h n\nonumber\\
	&&+2\epsilon^{a~~d}_{~bc}\hat{J}e^\mu_aR_{\mu\nu}^{bc}e^\nu_h\chi\nonumber\\
	&&\hspace{-2cm}
	-4\eta^{da}\eta_{bf}\eta_{ce}\hat{J}e^\mu_aR_{\mu\nu}^{bc}e^\nu_hs^{ef}\\
	mk^d+\chi^d+\frac{1}{2}i\hat{D}^d\chi-B^dn \qquad &&\nonumber\\
	+i(\eta_{ef}\delta^d_g-\eta_{eg}\delta^d_f)B^es^{fg} -4\epsilon_{efg}^{~~~h}
	(\lambda^{efg}+\frac{1}{2}i\hat{D}^es^{fg}) &=& 0\\
	m\tau_h^{~d}+\int p_hp^d{\cal P}d^4p+\frac{1}{2}i\hat{D}^d\chi_h-B^dU_h
		\qquad &&\nonumber\\
	+i(\eta_{ef}\delta^d_g-\eta_{eg}\delta^d_f)B^e\lambda_h^{~fg}
	-4\epsilon_{efg}^{~~~d}(\int p_hp^e{\cal T}^{fg}d^4p+\frac{1}{2}i\hat{D}^e
	\lambda_h^{~fg})&=& 4\epsilon^{da}_{~~bc}\hat{J}e^\mu_a
		R_{\mu\nu}^{bc}e^\nu_h n \nonumber\\
	&&\hspace{-2cm}
	-8i(\delta^a_b\delta^d_c-\delta^a_c\delta^d_b)\hat{J}e^\mu_aR_{\mu\nu}^{bc}
	e^\nu_h\chi\\
	ms^{ef}-\frac{1}{4}\hat{D}^{[e}j^{f]}+\frac{1}{2}iB^{[e}k^{f]} &=& 0
	 \label{eq:spin}\\
	m\lambda_h^{~ef}+\frac{1}{2}i\int p_hp^{[e}{\cal V}^{f]}d^4p -\frac{1}{4}
	\hat{D}^{[e}T_{|h|}^{~~f]}-\frac{1}{2}iB^{[e}\tau_{|h|}^{~~f]}&=&
	-12\epsilon^{a~~g}_{~bc}\epsilon^{ef}_{~~dg}\hat{J}e^\mu_aR_{\mu\nu}^{bc}
	e^\nu_h s^{dg}\nonumber\\
	&&\hspace{-2cm}
	+(\delta^a_c\epsilon^{ef}_{~~db}-\delta^a_d\epsilon^{ef}_{~~cb})\hat{J}
	e^\mu_aR_{\mu\nu}^{bc}e^\nu_hk^d
\end{eqnarray}
where $T=T_a^a, \tau = \tau_a^a$. Some of these have direct physical interpretations,
e.g., as we saw in the previous section (\ref{eq:confan}) gives a kinetic expression
for the conformal anomaly when $m=0$. In analogy with this, we will refer to the
equation for $\tau$, (\ref{eq:psconfan}), as the pseudo-conformal anomaly for lack of a
better word. Another important equation is (\ref{eq:heat}) which states that the momentum
density and the mass times the current density (which is also the velocity density)
are not identical, the difference between the two is an indication of heat flow.
Also, equation (\ref{eq:spin}) gives an expression for the spin density in terms of the
velocity/current density $j_a$ and the axial current $k_a$.\\
The quantities without a direct physical interpretation such as $\tau_{ab},\chi,\chi_a,
\lambda_{abc}$ can be eliminated from these equations.\\
One should also note that we are not able to eliminate the second moments appearing
in this set of equations. In general one will get an infinite hierarchy of moment
equations. This can be truncated by brute force at any given stage resulting in a set
of approximate hydrodynamic equations. The second moments have a physical interpretation
in terms of viscous pressure, since
\begin{equation}
	\Pi_h^{~d} := {\rm Tr}~\int p_hp^d\hat{W} d^4p =\int p_hp^d {\cal S} d^4p
\end{equation}
is the viscous pressure tensor, \cite{van}. The remaining second moments are then higher
Clifford algebra analogues of this, lacking a straightforward classical interpretation.

\section*{Arbitrary Spins}
Clearly the fermions are the most difficult to treat due to their Grassmannian
nature and that is why we have chosen to consider such fields in great detail.
Arbitrary spins are not that difficult. Consider a field equation
for a field $\Phi$ of some spin $s$,
\begin{displaymath}
    (D_s+M_s)\Phi = gJ
\end{displaymath}
where $D_s$ is some differential operator (first order whenever $s$ is half
integral and second order when $s$ is an integer), $M_s$ is a mass-term
including couplings to curvature such as $\xi R$ for $s=0$ and $R_{\mu\nu}$ for $s=1$,
$g$ is
a coupling constant and $J$ is a source term, as we have seen above such an
equation gives rise to an equation for the associated Wigner function $W_\Phi$
by the ``minimal substitution''
\begin{equation}
    i\nabla_\mu \mapsto e^a_\mu p_a+\frac{1}{2}i\nabla_\mu
\end{equation}
and the ``source term transformation''
\begin{equation}
    gJ\mapsto \hat{X}^{(s)}W_\Phi
\end{equation}
where $\hat{X}$ is some integro-differential operator containing the source.
For bosons with spin $s$, $\hat{X}$ is a $4s$ tensor and $W$ is a $2s$ tensor,
for $s=1/2$ we have already seen that $\hat{X}$ carries a Clifford algebra index
(i.e. two spinor indices) and three Lorentz indices, two of which must be understood
as coming from the Lie algebra $so(r,d-r)$, for $s=3/2$ we would get a Wigner function
which had two more Lorentz indices, and $\hat{X}$ would also have four extra Lorentz
indices to account for the vector index on the Rarita-Schwinger field. It is clear
that for $s\geq 1$ the notation quickly becomes cumbersome.\\
All Wigner functions are maps from the cotangent bundle into some algebra,
for a scalar field the target space is simply $\C$, for spin-$\frac{1}{2}$, 
it is $C(r,d-r)$, for $s=1$ we get $(T^*M\otimes \Lieg)\otimes(T^*M\otimes
\Lieg^\dagger)$ with $\Lieg$ the internal gauge algebra and so on. In general fermions
will have Wigner functions with values in $C(r,d-r)\otimes (T^*M)^{\otimes
(2s-1)}$ while bosonic fields will have Wigner functions taking their values
in $(T^*M)^{\otimes 2s}$, gauge
degrees of freedom are handled by simply enlarging the target space by 
tensoring it with $\Lieg\otimes\Lieg^\dagger$.\\
A spin-0 field, $\phi$, would then give rise to a Wigner equation of the form
\begin{equation}
    \left[\left(ie_\mu^a p_a+\frac{1}{2}i\nabla_\mu\right)^2+m^2+\xi R
	\right] W_\phi = \hat{X}^{(0)}W_\phi
\end{equation}
while the Maxwell field $A_\mu$ gives rise to
\begin{equation}
    \left(ie^a_\nu p_a-\frac{1}{2}\nabla_\nu\right)^2W_{\mu\rho}+R_\mu^{~\nu}
	W_{\nu\rho} =\hat{X}^{(1)~\nu\lambda}_{\mu\rho} W_{\nu\lambda}
\end{equation}
and a Rarita-Schwinger field $\psi_\mu$ would give rise to
\begin{equation}
    \frac{1}{2}\epsilon^{abcd}\gamma_5\gamma_be^\mu_ce^\nu_d\left(e^e_\mu p_e
    +\frac{1}{2}i\nabla_\mu\right)W_{\nu\rho} = \hat{X}^{(3/2)~\nu\lambda}_{\mu\rho} 
	W_{\nu\lambda}
\end{equation}
where we have omitted the $A$ and $\psi$ subscripts on the Wigner functions for
the cases of $s=1,3/2$ and suppresses the spinor indices in the latter case.\\
Yang-Mills fields could be treated analogously but will have more complicated
kinetic equations due to the non-linearity of their field equations.\\
Internal degrees of freedom is also treated easily. The fields will now be
cross sections in some associated bundle, and the parallel transporter has
to include not just the curvature effect but also the connection in the
corresponding principal bundle (the gauge field). One simply replaces the
covariant derivative with the appropriate gauge covariant derivative containing
the gauge field and the metric connection. If the field transforms in the 
representation $\rho$ of $\Lieg$ then the Wigner function will transform in the
representation $\bar{\rho}\otimes\rho$, if $\rho$ is the fundamental 
representation, then this is the adjoint one.

\subsection*{The Gravitational Field: Palatini Formalism}  
Next we write down the equations for the gravitational field. This can be viewed
from two points: (1) as a means of calculating the back-reaction of the quantum
fields on the space-time geometry (keeping the energy-momentum tensor fixed), or
(2) as constituting full-fledged quantum gravity (giving a set of coupled
equations).\\
We immediately face a problem: all-though the connection is the formal analogue
of the Yang-Mills fields, we cannot take over the results by Else, Gyulassy and
Vasak, as the Einstein equations are first order in the connection, while the
Yang-Mills equations are second order. In the presence of torsion the field
equations for gravity can be written
\begin{eqnarray}
    e^\nu_b R_{\mu\nu}^{ab} &=& \kappa(T_\mu^a+\frac{1}{2}e_\mu^a T)\\
    S^c_{ab} &=& 2\kappa\left(C_{ab}^c-\frac{3}{7}\delta^c_a C^d_{db}
    +\frac{2}{7}\delta^c_bC^d_{ad}\right)
\end{eqnarray}
where $T_\mu^a$ is the energy-momentum tensor, $T$ its trace, $S^a_{bc}$, the
torsion and $C^a_{bc}$ is given by
\begin{equation}
    C^c_{ab} \equiv \frac{1}{4}iE\bar{\psi}\gamma^c\sigma_{ab}\psi
\end{equation}
with $E=\det e_\mu^a$, see Ramond \cite{Ramond}. 
We will have to introduce {\em two} Wigner operators, one for the vierbein and
one for the spin-connection:
\begin{eqnarray}
    \Gamma^{ab;cd}_{\mu\nu} &\equiv & \int e^{-ip\cdot
    y}U_+\omega_\mu^{ab}\otimes U_-\omega_\nu^{cd} \frac{d^4y}{(2\pi)^4}\\
    L^{ab}_{\mu\nu} &\equiv & \int e^{-ip\cdot y}U_+ e^a_\mu\otimes U_- e^b_\nu
    \frac{d^4y}{(2\pi)^4}
\end{eqnarray}
Unfortunately these are non-covariant, as the spin connection  
transforms in an affine way under local $so(3,1)$-transformations. This 
non-covariance will then introduce an unwanted dependency on coordinate choices
(gauge dependency).\\ 
Finding the Wigner equations for these is straight-forward (albeit
tedious), and will not be done here, they will have the form
\begin{eqnarray}
    \left(e^a_\mu p_a+\frac{1}{2}i\nabla_\mu\right)\Gamma_{\nu\rho}^{ab;cd} -
    (\mu\leftrightarrow\nu) &=& \hat{X}_{\mu~gh}^{(T)~ab}\Gamma_{\nu\rho}^{gh;cd}-
    (\mu\leftrightarrow\nu)\\
    \left(e^a_\mu p_a+\frac{1}{2}i\nabla_\mu\right)L_{\nu\rho}^{ab} -
    (\mu\leftrightarrow\nu) &=& \hat{X}_{\mu~c}^{(C)~a}L_{\nu\rho}^{cb}-
    (\mu\leftrightarrow\nu)
\end{eqnarray}
Where $\hat{X}_\mu^{(T,C)}$ are integral operators containing the source terms.
Expressing these sources in terms of the Wigner operator for the matter fields,
e.g.\
\begin{equation}    
    T_{ab} \equiv  <{\rm Tr}(\int \gamma_a p_b\hat{W} d^4p)>
\end{equation}
would then lead to a set of coupled integro-differential equations, constituting
the full set of equations for quantum gravity. It is to be remembered that for
$g_{\mu\nu}$ a quantum field, we cannot take the curvature two-form out of the
integral in the original expression for $\hat{X}$. Therefore, the equation for
the matter Wigner function becomes much more complicated, in \cite{EGV} the 
analogous situation in \small{QCD} is treated. We will not attempt this 
level of generality here, just note that the straightforward generalisation 
from Yang-Mills fields to gravitational fields does not give tractable kinetic 
equations, at least not in the Palatini formalism.

\section*{The Gravitational Field: Ashtekar Variables}
As Elze, Gyulassy and Vasak, \cite{EGV}, have developed the Wigner function 
technique for Yang-Mills fields, one would suspect that an approach similar 
to theirs can be fruitful if one uses the Ashtekar formulation of gravity. 
In this formulation, \cite{Ashtekar}, one has a complex $SU(2)$-connection, 
$A=A_i^adx^i \sigma_a$, where $\sigma_a$ are the generators of $su_2$ and 
$i=1,2,3$ is a spatial index. This formalism is based on the fact that the 
Lorentz algebra $so(3,1)$ in four dimensions (and four dimension only) is 
isomorphic to the complexification of $su_2$. The Ashtekar formalism 
therefore makes use of two vital aspects of general relativity: the correct
dimensionality and the correct signature of spacetime. Canonically conjugate 
to the connection, we have the ``electric field'', $E^i_a$, i.e.
\begin{equation}
	\{A_i^a(x),E_b^j(x')\}_{PB} = \delta_b^a\delta_i^j\delta(x,x')
\end{equation}
Introducing the field strength tensor $F_{ij}^a$, the constraints can be 
written as
\begin{eqnarray*}
	D_i &=& E_a^jF_{ij}^a = 0\\
	H &=& \epsilon^{ab}_{~~c}E_a^iE_b^jF_{ij}^c = 0\\
	G_a &=& \nabla_jE^j_a = 0
\end{eqnarray*}
In exact analogy with the Yang-Mills case as presented in \cite{EGV}, we could
define
\begin{displaymath}
	\Gamma_{\mu\nu}^{ab}(x,p) = \int_{T_xM}\frac{d^4y}{(2\pi)^4)}
	e^{-ip\cdot y}
	\left(e^{-\frac{1}{2}y\cdot\nabla}F_{\mu\rho}^a\right)\otimes
	\left(F_{\sigma\nu}^{b\dagger} e^{\frac{1}{2}y\cdot\nabla^\dagger}
	\right) g^{\rho\sigma}
\end{displaymath}
with $F_{0i}^a = E_i^a$. Considering the Hamiltonian nature of the system, it 
is, however, more appropriate to introduce two slightly different 
Wigner functions (i.e. essentially splitting the above candidate into two), namely
\begin{eqnarray}
	\Gamma^{ab}_{ijkl} &=&\int_{T_x\Sigma}\frac{d^3y}{(2\pi)^3}e^{-ip\cdot y}
	\left(e^{-\frac{1}{2}y\cdot\nabla}F_{ij}^a\right)\otimes
	\left(F_{kl}^{b\dagger}e^{\frac{1}{2}y\cdot\nabla^\dagger}\right)\\
	L_{ab}^{ij} &=& \int_{T_x\Sigma}\frac{d^3y}{(2\pi)^3} e^{-ip\cdot y}
	\left(e^{-\frac{1}{2}y\cdot\nabla}E_a^i\right)\otimes
	\left(E_b^{j\dagger}e^{\frac{1}{2}y\cdot\nabla^\dagger}\right)
\end{eqnarray}
where we have used the global hyperbolicity of spacetime always assumed in a 
Hamiltonian formulation of gravity, $M\simeq \Sigma\times \R$, and furthermore
made explicit use of the complex nature of the connection and the ``electric 
field'' (whence the daggers on the right hand side).\\
The constraints can then be used to derive relationships between these two 
Wigner functions. The diffeomorphism constraint $D_i$ can be rewritten
\begin{equation}
	(E_a^i\otimes E_b^{j\dagger})(F_{ik}^a\otimes F_{jl}^{b\dagger})=0
\end{equation}
Noting that the Wigner functions $\Gamma_{ijkl}^{ab}, L_{ab}^{ij}$ are precisely
the Weyl transforms of these two tensor products, we see that the quantum version
of this constraint reads
\begin{equation}
	{\cal D}_{kl} \equiv L_{ab}^{ij}*\Gamma_{ikjl}^{ab} = 0
\end{equation}
where $*$ denotes the twisted product (the non-commutative product induced on the
set of phase-space functions by the non-Abelian product of operators),
\begin{equation}
	f*g \sim f e^{\frac{1}{2}i\stackrel{\leftrightarrow}{\bigtriangleup}} g
	= fg+\frac{1}{2}if\stackrel{\leftrightarrow}{\bigtriangleup} g...
\end{equation}
with $f\stackrel{\leftrightarrow}{\bigtriangleup}g := \{f,g\}$ being the Poisson
bracket.\\
Similarly the Hamiltonian constraint $H$ becomes
\begin{equation}
	{\cal H} \equiv \epsilon^{ab}_{~~c}\epsilon^{a'b'}_{~~~c'} 
	L_{aa'}^{ii'}*L_{bb'}^{jj'}*\Gamma_{iji'j'}^{cc'} = 0
\end{equation}
One should note that while the classical constraints $D_i, H$ are purely algebraic
in $F_{ij}^a, E_i^a$, the quantum versions become infinite order differential equations
in the corresponding Wigner functions -- equivalently, they become $*$-algebraic, i.e.
deformed.
The last constraint, the Gauss one, however, gives rise to a differential equation
for $L_{ij}^{ab}$, analogous to the equation for the Dirac-Wigner function (or more
precisely, to the Yang-Mills Wigner function as derived by Elze, Gyulassy and Vasak,
cite{EGV}). 

\section*{Group Theoretical Arguments}
We can round off this discussion with a few comments about the algebraic structure
behind the entire \small{WWM}-formalism.
The flat space Wigner function has a natural group theoretical interpretation,
\cite{group}. Define the operator
\begin{equation}
	\Pi(u,v) = \exp(iu\hat{p}-iv\hat{q})
\end{equation}
then $\Pi(u,v)$ form a ray representation of the (Abelian) group of translations in
phase-space,
\begin{equation}
	\Pi(u,v)\Pi(u',v') = e^{-i(uv'-vu')} \Pi(u+u',v+v')
\end{equation}
This operator then gives the Weyl transformation, mapping an operator into a
function on the classical phase-space
\begin{equation}
	A_W(u,v) = {\rm Tr}~\Pi(u,v)\hat{A}
\end{equation}
The Wigner function is the (symplectic) Fourier transform of the Weyl 
transform of the
projection operator (for a pure state) $|\psi\rangle\langle\psi|$, i.e. 
\begin{equation}
	W(x,p) = \int e^{i(up-vx)}\langle\psi|\Pi(u,v)|\psi\rangle dudv
\end{equation}
What we have done in this paper is to replace the Abelian group of translation
on a flat phase-space with the non-Abelian group of parallel transport on the
curved phase-space $T^*M$. This splits up into two parts, the parallel transport
in the base manifold $M$, which is generated by the momentum operator, and the
(Abelian) translations in the fibre $T^*_xM$, which is generated by 
$\hat{q}$.\\
In a previous work, \cite{wlie}, I have shown that one can generalise $\Pi$ to
a very large class of algebras, most notably finite dimensional Lie algebras,
their corresponding loop and Kac-Moody algebras as well as super Lie algebras 
and $C^*$-algebras. To put it in algebraic language, then, what one does when 
going from flat space to curved space is to replace the usual Heisenberg 
algebra by the curved space analogue
\begin{equation}
	[\hat{p}_\mu,\hat{p}_\nu] = -\hat{R}_{\mu\nu} \qquad
	[\hat{p}_\mu,\hat{q}^\nu] = -i\delta_\mu^\nu \qquad
	[\hat{q}^\mu,\hat{q}^\nu] = 0
\end{equation}
where $\hat{R}_{\mu\nu}$ is the curvature two-form ({\em not} the Ricci 
tensor).\\
The twisted product is given by
\begin{equation}
	A_W*B_W := (\hat{A}\hat{B})_W = {\rm Tr}~\Pi(u,v)\hat{A}\hat{B}
\end{equation}
and can be written in terms of a kernel as \cite{Kasperkovitz,wlie}
\begin{equation}
	(f*g)(u,v) = \int K(u,v,u',v',u'',v'')f(u',v')g(u'',v'') 
		du'dv' du''dv''
\end{equation}
with
\begin{equation}
	K(u,v,u',v',u'',v''):= {\rm Tr}~\Pi(u,v)\Pi(u',v')\Pi(u'',v'')
\end{equation}
Thus the quantity $\Pi$ is the essential ingredient in any generalised 
``Wigner-Weyl-Moyal formalism'' or ``symbol calculus'', \cite{wlie}. Hence what
is needed in general is (i) a ``symbol map'' sending operators (typically
pseudo-differential operators) into functions on the cotangent bundle, and (ii)
a map connecting two fibres, this latter map is simply the symbol of the
parallel translator. In principle one could have a different map in (ii), but 
the symbol of the parallel translator is the simplest choice.\\
The formula we have derived for $W$ was based on the phase-space of a classical
{\em mechanical} system, namely $T^*M$, the Wigner function took its values in
$C(3,1)\otimes \Gamma(T^*M)$ then, where $C(3,1)$ is the Clifford algebra and
$\Gamma(T^*M)$ denotes the set of cross sections of the cotangent bundle.
We used the group of parallel transport
and translations along the fibres to generalise the Wigner function from
quantum mechanics in flat spacetime. One can proceed to quantum field theory by
means of second quantisation, although this is not usually easy to define in a
curved background, \cite{QFT}. Locally this always makes sense, but there are 
often
global obstructions. Another method, which is not very common, is to treat the
field and its conjugate momenta as the fundamental phase-space (which is then
infinite dimensional), the Wigner function is then a functional of these 
fields. This is very similar to the way one treats quantum gravity in quantum 
cosmology, where one then considers the wave function of the universe. The 
Wigner function has already been extended to this situation. \\
Let us consider a field theory in Hamiltonian formalism and denote the fields
and their conjugate momenta by $\phi,\pi$ respectively. These can be either 
bosonic or fermionic. We then want a functional $\cal W$ such that
\begin{equation}
	\langle A\rangle \equiv \int {\cal W}[\phi,\pi] A[\phi,\pi]
	{\cal D}\phi{\cal D}\pi
\end{equation}
which generalises
\begin{displaymath}
	\langle \hat{A}\rangle = {\rm Tr}~\rho\hat{A} = \int A_W(x,p) 
		W(x,p) dxdp
\end{displaymath}
If we have free fields, then the phase space is flat, and the parallel 
transporter becomes simply
\begin{equation}
	\Pi = e^{i\int u\hat{\phi}-v\hat{\pi} dx}
\end{equation}
From the quantum mechanical relation
\begin{displaymath}
	W(u,v) = {\rm Tr}~\rho\Pi(u,v) = \langle \Pi(u,v)\rangle
\end{displaymath}
which is the symplectic Fourier transform of the Wigner function, we conclude
\begin{equation}
	{\cal W}(u,v) := \langle \Pi(u,v)\rangle = \frac{1}{Z(0)}\int e^{iS}
	\Pi(u,v){\cal D}\phi{\cal D}\pi
\end{equation}
Noticing the concrete form of $\Pi$ we see that $\exp(iS)\Pi$ is equivalent to
adding a source term (with the sources denoted by $u,v$ respectively), hence
\begin{equation}
	{\cal W}(u,v) = \frac{Z(u,v)}{Z(0)}
\end{equation}
The proper Wigner functional is then a symplectic (functional) Fourier 
transform of this quantity. This makes it possible to interpret the partition 
function (with sources) as essentially the Wigner functional of the vacuum 
state (or more general, the vacuum to vacuum transition).

\section*{Conclusion}
We have generalised the work by Elze, Gyulassy and Vasak to gravitation and
quantum fields in curved space-time. Thereby we obtained a set of {\em exact}
equations for QFT in curved space-time and even Quantum Gravity, which 
allows us
to make {\em non-perturbative} calculations in these cases. The major draw-back
is the complicated nature of the equations -- especially in the full quantum
gravity case. But on the
other hand, we can develop a recursive scheme for these quantum corrections
(the $\bigtriangleup$-expansion), and
we can therefore avoid perturbation theory all together. One should also note
that this approach is on the one hand intimately related to the algebraic structure
of the space of quantum observables (the canonical commutator relations for the
fields), and on the other to the topological structure of spacetime (the
Wigner function is, for Dirac fermions, a mapping from the cotangent bundle
into the Clifford algebra, whereas for bosons it is a mapping from the
cotangent bundle into a the tensor algebra). It was this interplay that
allowed us to find dynamical expressions for the conformal anomaly in $d=2$
and $d=4$.\\
We also saw that the kinetic equation satisfied by the Wigner function could
be split up into two, one being the mass-shell constraint giving the quantum
and curvature induced corrections to the mass, while the other was the kinetic
equation proper. In certain classical-like situations these could be
written as two Fokker-Planck equations, the one with no momentum diffusion
the other with no dynamical friction. We derived expressions for these
kinetic quantities. We also saw how quantum corrections modified this simple
situation.\\
By taking appropriate moments of the Wigner equation we arrived at a set of coupled
equations governing macroscopic quantities such as energy-momentum tensor and
current and spin densities. These were the corresponding hydrodynamic equations. We saw
that they, besides giving the kinetic interpretation of the conformal anomaly, also
lead to an expression for the heat flow.\\
Gravitational degrees of freedom was attempted handled first in the Palatini
formalism, in which we had to introduce Wigner functions for as well the
vierbein as the spin connection. These was not, however, covariant, and the
resulting equations were too complicated. We then turned to the Ashtekar
variables, where we could either introduce one Wigner function for 
$F_{\mu\nu}^a$ in the standard way, or use a canonical description to split
this into two, one for $F_{ij}^a$ and the other for $E_a^i$. These were by
construction covariant. The constraints induced conditions on these Wigner functions,
which were $*$-algebraic, i.e. infinite order differential relations due to a
quantum deformation of the product.

\end{document}